\documentclass[10pt,final,journal,compsoc]{IEEEtran}
\usepackage[utf8]{inputenc}
\usepackage{graphicx}
\usepackage{amsmath}
\usepackage[hidelinks]{hyperref}
\usepackage{tabularx}
\usepackage{caption}
\usepackage{booktabs}
\usepackage{rotating}
\usepackage{multirow}
\usepackage{multicol}
\usepackage[nobreak,space,compress]{cite}

\usepackage{wrapfig}
\usepackage{ragged2e}
\usepackage{comment}
\usepackage{array,makecell}
\usepackage{rotating}
\usepackage{arydshln}
\newcommand{\Flakify}{Flakify}

\usepackage{marginnote}

\usepackage[backgroundcolor=white,bordercolor=blue,linecolor=blue]{todonotes}
\makeatother
\begin{document}

\title{\Flakify: A Black-Box, Language Model-based Predictor for Flaky Tests}

\author{
    Sakina Fatima, 
    Taher A. Ghaleb, and
    Lionel Briand, \IEEEmembership{Fellow, IEEE}
\IEEEcompsocitemizethanks{
    \IEEEcompsocthanksitem S. Fatima is with the School of EECS, University of Ottawa, Ottawa, Canada.~E-mail: sfati077@uottawa.ca

    \IEEEcompsocthanksitem T. A. Ghaleb is with the School of EECS, University of Ottawa, Ottawa, Canada.~E-mail: tghaleb@uottawa.ca

    \IEEEcompsocthanksitem L. Briand holds shared appointments with the SnT Centre for Security, Reliability and Trust, University of Luxembourg, Luxembourg and the school of EECS, University of Ottawa, Ottawa, Canada.~E-mail: lbriand@uottawa.ca
}
\thanks{}}

\markboth{IEEE TRANSACTIONS ON SOFTWARE ENGINEERING}%
{Fatima \MakeLowercase{\textit{et al.}}: \Flakify:A Black-Box, Language Model-based Predictor for Flaky Tests}

\IEEEtitleabstractindextext{
    \begin{abstract}\justifying
        Software testing assures that code changes do not adversely affect existing functionality. However, a test case can be flaky, i.e., passing and failing across executions, even for the same version of the source code. Flaky test cases introduce overhead to software development as they can lead to unnecessary attempts to debug production or testing code. Besides rerunning test cases multiple times, which is time-consuming and computationally expensive, flaky test cases can be predicted using machine learning (ML) models, thus reducing the wasted cost of re-running and debugging these test cases. However, the state-of-the-art ML-based flaky test case predictors rely on pre-defined sets of features that are either project-specific, i.e., inapplicable to other projects, or require access to production code, which is not always available to software test engineers. Moreover, given the non-deterministic behavior of flaky test cases, it can be challenging to determine a complete set of features that could potentially be associated with test flakiness. Therefore, in this paper, we propose \Flakify, a black-box, language model-based predictor for flaky test cases. \Flakify~relies exclusively on the source code of test cases, thus not requiring to (a) access to production code (black-box), (b) rerun test cases, (c) pre-define features. To this end, we employed CodeBERT, a pre-trained language model, and fine-tuned it to predict flaky test cases using the source code of test cases. We evaluated \Flakify~on two publicly available datasets (FlakeFlagger and IDoFT) for flaky test cases and compared our technique with the FlakeFlagger approach, the best state-of-the-art ML-based, white-box predictor for flaky test cases, using two different evaluation procedures: (1) cross-validation and (2) per-project validation, i.e., prediction on new projects. \Flakify~achieved F1-scores of 79\% and 73\% on the FlakeFlagger dataset using cross-validation and per-project validation, respectively. Similarly, \Flakify~achieved F1-scores of 98\% and 89\% on the IDoFT dataset using the two validation procedures, respectively. Further, \Flakify~surpassed FlakeFlagger by 10 and 18 percentage points (pp) in terms of precision and recall, respectively, when evaluated on the FlakeFlagger dataset, thus reducing the cost bound to be wasted on unnecessarily debugging test cases and production code by the same percentages (corresponding to reduction rates of 25\% and 64\%). \Flakify~also achieved significantly higher prediction results when used to predict test cases on new projects, suggesting better generalizability over FlakeFlagger. Our results further show that a black-box version of FlakeFlagger is not a viable option for predicting flaky test cases.
    \end{abstract}

}

\maketitle
\IEEEdisplaynontitleabstractindextext
\IEEEpeerreviewmaketitle

\section{Introduction}
  Software testing is an essential activity to assure software dependability. When a test case fails, it usually indicates that recent code changes were incorrect. However, it has been observed, in many environments, that test cases can be non-deterministic, passing and failing across executions, even for the same version of the source code. These test cases are referred to as flaky test cases~\cite{zolfaghari2021root,luo2014empirical,eck2019understanding}. Flaky test cases can introduce overhead to software development, since they require developers to either (a) debug the production or testing code looking for a bug that might not really exist, or (b) rerun a failed test case multiple times to check if it would eventually pass, thus suggesting that the failure is not due to recent code changes but to the test case itself.

  Previous research has investigated the common reasons behind test flakiness, such as concurrency, resource leakage, and test smells. The conventional approach to detect flaky test cases is to rerun them numerous times~\cite{bell2018deflaker,lam2019idflakies}, which is in most practical cases computationally expensive~\cite{GoogleFlakyCost} or even impossible. To address this issue, recent studies have proposed approaches using machine learning (ML) models to predict flaky test cases without rerunning them~\cite{alshammari2021flakeflagger,pinto2020vocabulary,camara2021use}, thus proposing a much more scalable and practical solution. Despite significant progress, these approaches (a) rely on production code, which is not always accessible by software test engineers or a scalable solution, or (b) employ project-specific features as flaky test case predictors, which makes them inapplicable to other projects. Moreover, these approaches rely on a limited set of pre-defined features, extracted from the source code of test cases and the system under test. However, when evaluated on realistic datasets, these approaches yield a relatively low accuracy (F1-scores in the range 19\%-66\%), thus suggesting they may not capture enough information about test flakiness. Finding additional features that could potentially be associated with flaky test cases, preferably based on test code only (black-box), is therefore a research challenge. 
  
  In this paper, we propose \Flakify~(Flaky Test Classify), a generic language model-based solution for predicting flaky test cases. \Flakify~is black-box as it relies exclusively on the source code of test cases (test methods), thus not requiring access to the production code of the system under test. This is important as production code is not always (entirely) accessible to test engineers due, for example, to outsourcing software testing to a third-party. Further, analyzing production code may raise many scalability and practicality issues, especially when applied to large industrial systems using multiple programming languages. In addition, \Flakify~does not require the definition of features---which are necessarily incomplete---to be used as predictors for flaky test cases. Instead, we used CodeBERT~\cite{feng2020codebert}, a pre-trained language model, and fine-tuned it to classify test cases as flaky or not based on their source code. To improve \Flakify, we further pre-processed test code to remove potentially irrelevant information. We evaluated \Flakify~on two different datasets: the FlakeFlagger dataset, containing 21,661 test cases collected from 23 Java projects, and the IDoFT dataset, containing 3,862 test cases collected from 312 Java projects. To do this, we used two different evaluation procedures: (1) cross-validation and (2) per-project validation, i.e., prediction on new projects. Our results were compared to FlakeFlagger~\cite{alshammari2021flakeflagger}, the best state-of-the-art ML-based predictor for flaky test cases. Specifically, our evaluation addresses the following research questions.
  
  \begin{itemize}
      \item \textbf{RQ1: How accurately can \Flakify~predict flaky test cases?}\\
      \Flakify~achieved promising prediction results when evaluated using two different datasets. In particular, based on cross-validation, \Flakify~achieved a precision of 70\%, a recall of 90\%, and an F1-score of 79\% on the FlakeFlagger dataset, and a precision of 99\%, a recall of 96\%, and an F1-score of 98\% on the IDoFT dataset. \Flakify~yielded slightly worse results when predicting flaky tests on new projects, with a precision of 72\%, a recall of 85\%, and an F1-score of 73\% on the FlakeFlagger dataset, and a precision of 91\%, a recall of 88\%, and an F1-score of 89\% on the IDoFT dataset.
      
      \vspace{3pt}
      \item \textbf{RQ2: How does \Flakify~compare to the state-of-the-art predictors for flaky test cases?}\\
      The best performing model of \Flakify~achieved a significantly higher precision (70\% vs. 60\%) and recall (90\% vs 72\%) on the FlakeFlagger dataset in predicting flaky test cases than FlakeFlagger, the best state-of-the-art, white-box approach for predicting flaky test cases. Hence, with \Flakify, the cost of debugging test cases and production code is reduced by 10 and 18 percentage points (pp) (a reduction rate of 25\% and 64\%), respectively, when compared to FlakeFlagger. Moreover, our results show that a black-box version of FlakeFlagger is not a viable option for predicting flaky test cases. Specifically, FlakeFlagger became 39 pp less precise with 20 pp less recall when only black-box features were used as predictors for flaky test cases. 
      
      \vspace{3pt}
      \item \textbf{\textit{RQ3: How does test case pre-processing improve \Flakify?}}\\
      Retaining only code statements that are related to a selected set of test smells improved the precision, recall, and F1-score of \Flakify~by 5 pp and 6 pp on the FlakeFlagger and IDoFT datasets, respectively. The goal was to address a limitation of CodeBERT (and all other language models), which leads to only considering the first 512 tokens in the test source code. This result also confirms the previously reported association of test smells with flaky test cases~\cite{palomba2017notice,alshammari2021flakeflagger,camara2021use,pontillo2021toward}.
  \end{itemize}
  
  Overall, this paper makes the following contributions.
  \begin{itemize}
      \item A generic, black-box, language model-based flaky test case predictor, which does not require rerunning test cases.
      \item An ML-based classifier that predicts flaky test cases on the basis of test code without requiring the definition of features.
      \item An Abstract Syntax Tree (AST)-based technique for statically detecting and only retaining statements that match eight test smells in the test code, thus enhancing the application of language models.
  \end{itemize}

   The rest of this paper is organized as follows.
   Section~\ref{background} provides background about flaky test cases and language models.
   Section~\ref{approach} presents our black-box approach for predicting flaky test cases.
   Section~\ref{validation} evaluates our approach, reports experimental results, and discusses the implications of our research..
   Section~\ref{threats} discusses the validity threats to our results. 
   Section~\ref{related_work} reviews and contrasts related work.
   Finally, Section~\ref{conclusion} concludes the paper and suggests future work. 

\section{Background}
\label{background}
In this section, we describe flaky test cases, their root causes, their practical impact, and the strategies to detect them. In addition, we describe pre-trained language models and how they can potentially contribute to predicting flaky test cases.

    \subsection{Flaky Test Cases}
    In software testing, a flaky test refers to test cases that intermittently fail and pass across executions, even for the same version of the source code, i.e., non-deterministically behaving  test cases~\cite{zolfaghari2021root}. Flaky test cases lead to many problems during software testing, by producing unreliable results and wasting time and computational resources. A flaky test can also fail for different reasons across executions, making it difficult to identify which failures are actually related to faults in the system under test.

    Flaky test cases have been reported to be a significant problem in practice at many companies including Google, Huawei, Microsoft, SAP, Spotify, Mozilla, and Facebook~\cite{ziftci2020flake,lam2019root,bach2017coverage,haben2021replication}. As reported by Google, almost 16\% of their 4.2 million test cases are flaky~\cite{GoogleFlakyCost}. Microsoft has also reported that 26\% of $3.8k$ build failures were due to flaky test cases. Many studies have been conducted to study flaky test cases, their causes, and the solutions to address them~\cite{zolfaghari2021root, parry2021survey,luo2014empirical,alshammari2021flakeflagger,bell2018deflaker,pinto2020vocabulary,camara2021use,pontillo2021toward}. Prominent causes of flaky test cases include asynchronous waits, test order dependency, concurrency, resource leakage, and incorrect test inputs or outputs. In addition, flaky test cases were found to be associated with other factors, such as test smells, which are further discussed below.

    \subsection{Flaky Test Case Detection}
        The most common approach for detecting flaky test cases is by rerunning test cases numerous times to check whether they behave consistently across executions~\cite{bell2018deflaker,lam2019idflakies}. Though effective, this approach is computationally expensive and not practical in many situations, for example in continuous integration contexts, where builds are submitted automatically and frequently to perform regression testing. To mitigate such cost, other approaches attempted to detect flaky test cases without relying on rerunning them. To that end, characteristics of test cases, such as execution history, coverage information, and static test features, were used to predict whether a test case is flaky or not. Prediction models were built using ML and Natural Language Processing (NLP) techniques~\cite{alshammari2021flakeflagger,pinto2020vocabulary,camara2021use}. Such techniques require training ML models with pre-defined sets of features used as indicators for test flakiness. Such features commonly present practical limitations, such as (a) their reliance on production code, which is not always accessible or efficiently analyzable by test engineers, and (b) their limited capacity to capture the actual structure or behavior of test cases, such as the use of language keywords~\cite{pinto2020vocabulary} or the presence of test smells~\cite{alshammari2021flakeflagger,camara2021use,pontillo2021toward} in test code.
        
        After identifying potentially flaky test cases, developers can focus their investigation on them and, hence, attempt to fix code statements causing such flakiness. Developers may also choose to rerun those specific test cases many more times to verify that they are actually flaky~\cite{parry2022surveying}. This is a reasonable undertaking, since test cases predicted as flaky normally represent a small percentage of the entire test suite. This, in turn, significantly eliminates a large part of the effort and time required to investigate or rerun test cases whenever a failure occurs~\cite{alshammari2021flakeflagger}.

    \subsection{Test Smells}
    Test Smells are inappropriate design or implementation choices made by developers while writing test cases~\cite{van2001refactoring}. Though test smells might not harm the functionality of a test case, previous research has reported that they tend to be associated with test flakiness. Palomba et al.~\cite{palomba2017notice} reported that nearly two-thirds of flaky test cases were associated with at least one test smell. Test smells were further employed to classify whether a test case is flaky or not. For example, test smells in Table~\ref{tab:TestSmells} were part of the features used by Alshammari et al.~\cite{alshammari2021flakeflagger} to predict test flakiness. Camara et al.~\cite{camara2021use} also used a more comprehensive set of test smells for flaky test case prediction. Results showed that \textit{Sleepy Test} and \textit{Assertion Roulette} are among test smells that are highly associated with flaky test cases.

     \begin{table*}
        \centering
        \caption{Test smells used by FlakeFlagger~\cite{alshammari2021flakeflagger}}
        \vspace{-7pt}
        \begin{tabular}{lll}
        \toprule
            \textbf{Test Smell} & \textbf{Description}\\
        \midrule
            Indirect Testing  & A test interacts with the class under test using methods from other classes\\
            Eager Testing  & A test performs multiple checks for various functionalities \\
            Test Run War &  A test allocates files or resources that might be used by other test cases\\
            Conditional Logic & A test uses a conditional \textit{if} statement\\
            Fire and Forget & A test launches background threads or processes \\
            Mystery Guest & A test accesses external resources \\
            Assertion Roulette & A test performs multiple assertions \\
            Resource Optimism & A test accesses external resources without checking their existence \\
        \bottomrule
        \end{tabular}
        \label{tab:TestSmells}
        \vspace{-7pt}
    \end{table*}
    
    \subsection{Pre-trained Language Models}
        Much research has been carried out in the field of NLP for developing pre-trained language models. Language models estimate the probability of different linguistic units, i.e., words, symbols, and sequence of them, occurring in a given sentence. There are many language models proposed in the literature, such as BERT~\cite{devlin2018bert}, ELMo~\cite{peters2018deep}, XLNet~\cite{yang2019xlnet}, RoBERTa~\cite{liu2019roberta}, and VideoBERT~\cite{sun2019videobert}. These models were pre-trained, using self-supervised learning, on a large corpus of unlabelled data. For example, BERT was pre-trained using a large dataset of English text collected from books and Wikipedia, whereas VideoBERT was pre-trained using a large dataset of instructional videos collected from YouTube.
        
        Pre-trained language models are often further fine-tuned using a specific, labelled dataset to train neural networks for performing various NLP tasks, such as text classification and entity recognition~\cite{nadeau2007survey}, relation extraction~\cite{bach2007review},  sentence tagging, or next sentence prediction~\cite{devlin2018bert}. For example, BERT was fine-tuned to perform sentiment analysis~\cite{xu2019bert,sun2019fine}, trained on labelled datasets to assign sentiment tags, i.e., positive, negative, or neutral, to a given text. Fine-tuning requires initializing a language model with the same parameters used for pre-training, and then further training the model using labeled data related to a specific task.

        Language models usually employ multi-layer transformers as a model architecture to perform many computations in parallel~\cite{vaswani2017attention}. Transformer models adopt positional embedding to vectorize individual words by considering their positions in a given sequence of words. Thus, unlike Recurrent Neural Networks (RNNS)~\cite{mandic2001recurrent} and Long-Short Term Memory (LSTM)~\cite{schmidhuber1997long}, transformer models do not require looking at past hidden states to capture dependencies with previous words in a sequence of words.
    
        Given the wide popularity of language models in various NLP applications, researchers have attempted to apply these language models to programming languages. However, when BERT, for example, was used for detecting the architectural tactics in source code~\cite{keim2020does}, e.g., recognizing software design patterns, the results were relatively worse compared to those obtained when BERT was used for natural language text. To address this issue, recent work proposed pre-training language models on source code written in many programming languages in addition to natural language text~\cite{feng2020codebert,guo2020graphcodebert,kanade2020learning,jiang2021treebert}. These models are well suited for fine-tuning to perform tasks related to source code. CodeBERT~\cite{feng2020codebert} is an example of a language model that was pre-trained on both natural and programming languages.
    
    \subsubsection{CodeBERT}
        CodeBERT~\cite{feng2020codebert} is a language model that was pre-trained on a large, unlabeled dataset containing English text as well as source code written in six different programming languages, namely Java, JavaScript, Python, Ruby, PHP, and Go, obtained from the CodeSearchNet corpus~\cite{husain2019codesearchnet}. CodeBERT takes, as input, source code statements and natural language sentences, which are then tokenized using the WordPiece~\cite{wu2016google} tokenizer. Similar to BERT and RoBERTa, CodeBERT uses a multi-layer bidirectional transformer~\cite{vaswani2017attention} as model architecture. This transformer is composed of six layers, each of which contains 12 self-attention heads capturing word relationships, a hidden state, and a 768-dimensional vector, as the output of each layer.
         
        CodeBERT also employed Masked Language Modeling (MLM)~\cite{devlin2018bert} and Replaced Token Detection (RTD)~\cite{clark2020electra} during pre-training, allowing to take tokens from random positions and masking them with special tokens, which are later used to predict the original tokens. As a result, each token is assigned a vector representation containing information about the token and its position in a given code. The final output of CodeBERT is a single vector representation aggregating all individual vector representations. This vector representation can further be fine-tuned to perform various tasks, e.g., classification. For example, to evaluate the performance of CodeBERT, it was fine-tuned to perform two tasks: (1) code search, i.e., retrieving the most relevant code to a given natural language text; (2) code documentation, i.e., generating a natural language description for a given source code. Moreover, CodeBERT was also adopted to perform classification tasks, such as bug prediction~\cite{pan2021empirical} and vulnerability detection~\cite{wu2021literature}.

    \subsubsection{Other models for programming languages}
        As mentioned above, recently, many language models for programming languages were proposed. For example, GraphCodeBERT~\cite{guo2020graphcodebert} was pre-trained on the inherent structure of source code and its data flow showing variables dependencies. Similar to CodeBERT, GraphCodeBERT was used for code search, in addition to code translation and refinement as well as clone detection. Another model for programming languages is TreeBERT~\cite{jiang2021treebert}, which was pre-trained using AST representations of Java and Python source code. TreeBERT was used for code documentation, similar to CodeBERT, in addition to code summarization. There is also CuBERT~\cite{kanade2020learning}, a programming language model pre-trained using Python source code. CuBERT was used for classification tasks, such as classifying exceptions and variable misuses. 
        
        Despite the capabilities of these models, CodeBERT has been the most commonly used language model and we selected it to address our objectives for several reasons presented below.
        \begin{itemize}
            \vspace{-5pt}
            \item The pre-trained CodeBERT model is publicly available.\footnote{\url{https://huggingface.co/microsoft/CodeBERT-base}}
            
            \item Unlike GraphCodeBERT, CodeBERT does not take into consideration the data flow in a given source code, which might not be easy to capture using test code only. For example, unlike local variables, if a global or external variable is used by a test case, GraphCodeBERT cannot identify the type and value of that variable when analyzing test code only.

            \item Unlike TreeBERT, which requires converting source code into ASTs, CodeBERT only requires source code as input.
            
            \item Unlike CuBERT, which was only pre-trained on Python source code without comments, CodeBERT was pre-trained on multiple programming languages using both source code and natural language comments.
            \vspace{-5pt}
        \end{itemize}

\section{Black-Box Flaky Test Case Predictor}
\label{approach}
This section describes our black-box solution for predicting flaky test cases. This is motivated by making such predictions scalable, as white-box analysis of the production source code, especially in the context of large systems, is often not a viable solution. 

    \subsection{CodeBERT for Flaky Test Case Prediction}
    In this paper, we propose \Flakify, a black-box solution for predicting whether a test case is flaky or not. \Flakify~relies solely on the source code of a test case and does not require to rerun it multiple times. The source code of test cases, i.e., Java test methods, includes the method declaration, body, and it associated Javadoc comments. While several studies have proposed ML techniques to predict flaky test cases, such techniques rely on pre-defined features extracted not only from the source code of test cases but also that of the system under test. However, results~\cite{alshammari2021flakeflagger,pinto2020vocabulary,camara2021use} suggest those features may not be enough, and finding additional features that could potentially be associated with flaky test cases remains a research challenge given their non-deterministic behavior. Therefore, we employed CodeBERT, the pre-trained language model described above, to perform a binary classification of test cases as \textit{Flaky} or \textit{Non-Flaky}. CodeBERT does not require to define features as it automatically identifies patterns based on the syntax and semantics of a given test code.
    
    CodeBERT starts by converting the source code of a test case into a list of tokens, each of which is converted into an integer vector representation. Finally, an aggregated vector representation is generated as an output of CodeBERT, which is further fine-tuned to classify test cases as \textit{Flaky} or \textit{Non-Flaky}. Figure~\ref{fig:tokenization} presents an example of how the source code of a test case is converted into tokens and then into integer vector representations. 

    \subsubsection{Source Code Tokenization} 
    To transform the source code into tokens, the source code of test cases is tokenized by the WordPiece~\cite{wu2016google} tokenizer using a pre-generated vocabulary file containing the vocabulary of both English and programming languages used for model pre-training. However, uncommon words, i.e., those that do not exist in the vocabulary file, are separated into several sub-words. 
    For example, the CodeBERT tokenizer splits `\texttt{assertThat}' into `\texttt{assert}' and `\texttt{\#\#that}', where `\texttt{\#\#}' denotes that a token represents a sub-word. Then, if a token is not found in the vocabulary file, the unknown token, $<$\texttt{UNK}$>$, is used. For each input, two special tokens, \texttt{[CLS]} and \texttt{[SEP]}, are added. Eventually, for a given source code, the tokenizer generates a sequence of tokens in the form of \texttt{[CLS]}$, c_1, c_2, .., c_n,$ \texttt{[SEP]}, where $c_i$ is a code token. The \texttt{[CLS]} token plays an important role in the classification of flaky test cases, as it contains the aggregated vector representation of all the vector representations of the tokens of a given test case. On the basis of that aggregated vector representation, our model classifies a test case as \textit{Flaky} or \textit{Non-Flaky}. \texttt{[SEP]} is just used to mark the end of the sequence of tokens. The tokenizer also adds `\texttt{Ġ}' in front of each word that is preceded by a whitespace in a statement.

    \subsubsection{Converting Tokens into Vector Representations} 
\begin{figure}[ht]
        \centering
        \includegraphics[width=0.70\linewidth]{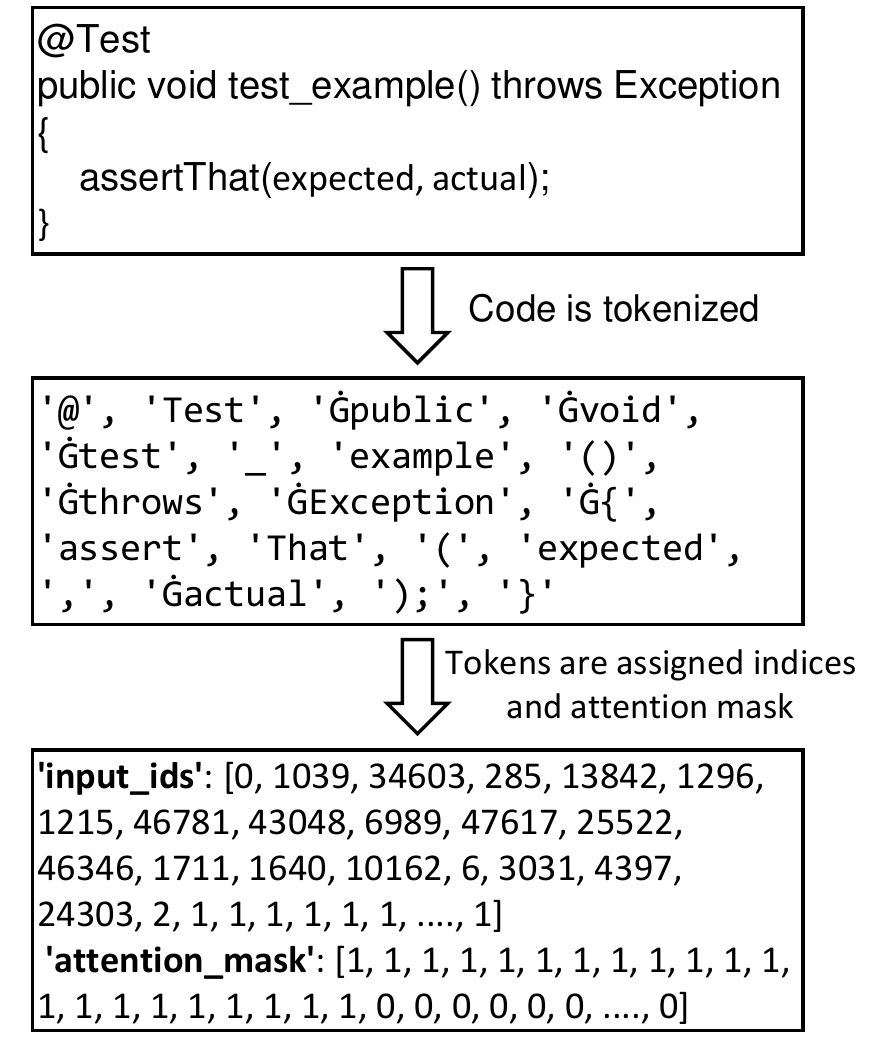}

        \caption{The process of converting the source code of a test case into a sequence of tokens, where each token is assigned an input index (id) and attention mask. Dots `....' are used to save space, since the actual length is 512. The input id of each token refers to a 768-dimensional vector representation.}
        \label{fig:tokenization}
        \vspace{-5pt}
    \end{figure}

    Once the source code tokens are generated, each token, including sub-word, special, and unknown tokens, is mapped to an index, e.g., id 34603 for ``Test" in Figure~\ref{fig:tokenization}, based on the position and context of each word in a given input. Each token is described by an 768-dimensional integer vector generated during CodeBERT pre-training. Using token padding, the same token length is given to the code of all test cases used as input, e.g., ``1" in Figure~\ref{fig:tokenization}.
    However, CodeBERT has a limit of 512 tokens per input. As a result, any token sequence exceeding that limit is truncated, which might lead to removing code statements with potentially relevant information about test flakiness. In addition to \textit{input ids} matching tokens, another list of \textit{attention masks} is generated containing ones and zeros to help the model distinguish between code tokens, which should be given attention, and extra tokens added for padding. Finally, for each test case, token vectors are aggregated to form one vector characterizing the \texttt{[CLS]} token, which is also represented using a 768-sized vector referred to by the first input index `0'.

    \subsubsection{Fine-Tuning CodeBERT for Flaky Test Classification}
\begin{figure}
        \centering
        \includegraphics[width=1\linewidth]{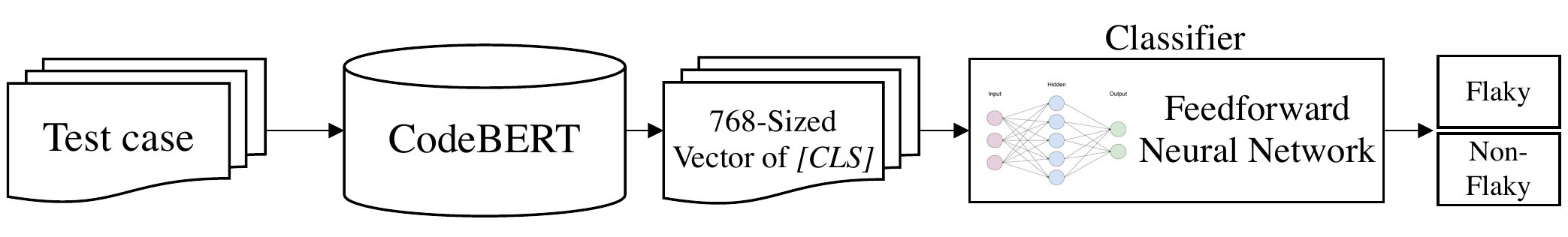}
        \caption{Fine-tuning CodeBERT for classifying test cases as flaky or not}
        \label{fig:fine_tuning}
        \vspace{-10pt}
    \end{figure}

    CodeBERT was pre-trained with a huge number of parameters, enabling it to recognize the source code structure. As a result, if CodeBERT were to be trained from scratch on our dataset, it would result into over-fitting. To avoid that, CodeBERT, similar to other language models~\cite{howard2018universal}, needs to be fine-tuned using data representative of the problem at hand. To do this, we employed CodeBERT as pre-trained and use its outputs, on our dataset, to train a Feedforward Neural Network (FNN)  to perform binary classification of test cases as flaky or non-flaky, as shown in Figure~\ref{fig:fine_tuning}.
    
    The output of CodeBERT, i.e., the aggregated vector representation of the \texttt{[CLS]} token, is then fed as input to a trained FNN to classify test cases as flaky or not. The FNN contains an \textit{input} layer of 768 neurons, a \textit{hidden} layer of 512 neurons, and an \textit{output} layer with two neurons. We used ReLU~\cite{agarap2018deep} as an activation function, which helps to speed up training by transforming the data within layers and output the \textit{input} directly if it is positive or zero otherwise. Then, we added a \textit{dropout} layer~\cite{srivastava2014dropout} to eliminate some neurons randomly from the network, by resetting their weights to zero during the training phase to prevent model over-fitting~\cite{el2021bert}. We used the \textit{Softmax} function to compute the probability of a test case to be \textit{Flaky} or \textit{Non-Flaky}. We used a learning rate of $10^{-5}$ using the AdamW optimizer~\cite{yao2020adahessian} and employed a batch size of two due to computational limitations. Using this configuration, we further trained CodeBERT on our training and validation datasets, which enabled the selection of improved parameter values for weights and biases through back propagation. We then evaluated the model, with the obtained weights, using a test dataset.
    
    \subsection{Identifying Test Smells}
    \label{Identifying_Test_Smells}
    As indicated above, the 512 token length limit induced by CodeBERT truncates longer test code, which leads to losing potentially relevant information about test flakiness. Therefore, we pre-possessed the source code of test cases to reduce their token length by only retaining information believed to be more relevant to test flakiness. To this end, for test cases exceeding the token length limit, we retained only code statements that match at least one of the eight test smells that were used by FlakeFlagger~\cite{alshammari2021flakeflagger} as predictors for flaky test cases. We also retained the method declaration and the associated Javadoc, since the signature and natural language description, if any, of the test case, might contain key terms or phrases that are likely associated with test flakiness, e.g., \textit{"...failures...unnecessary..."} or \textit{"thread-safe"}. 

    There exist several open source tools available for detecting test smells~\cite{aljedaani2021test}. However, these tools, e.g., \textit{tsDetect}~\cite{peruma2020tsdetect} and \textit{JNose Test}~\cite{virginio2020jnose}, either rely on production code for detecting test smells or do not detect all test smells that are potentially relevant to test flakiness~\cite{palomba2017notice,alshammari2021flakeflagger}. While Alshammari et al.~\cite{alshammari2021flakeflagger} detects all the eight test smells shown in Table~\ref{tab:TestSmells}, their technique does so by running test cases and requiring access to the production code for smell detection. Though we were inspired by the heuristics used by Alshammari et al. to detect test smells, given that our approach aims to be black-box, we developed an entirely different technique that detects test smells statically, relying exclusively on test code without requiring to run test cases. \Flakify~detects all targeted test smells and can be easily extended to detect additional test smells. We used an Abstract Syntax Tree (AST)~\cite{noonan1985algorithm} parser, provided by the Eclipse JDT library,\footnote{\url{https://www.eclipse.org/jdt}} to statically traverse any given test code and retain statements that match any of the targeted test smells.Using this library, each Java file in a test suite is parsed and converted into AST nodes representing different code elements, e.g., method declaration or invocation. Then, an AST visitor is used to traverse those AST nodes. We extended the AST visitor to check the AST nodes related to method declarations and apply heuristics (described below) to detect and retain code statements that match at least one test smell. Such statements are extracted as part of the pre-processed code. 

\begin{figure}[ht]
        \centering
        \includegraphics[width=1\linewidth]{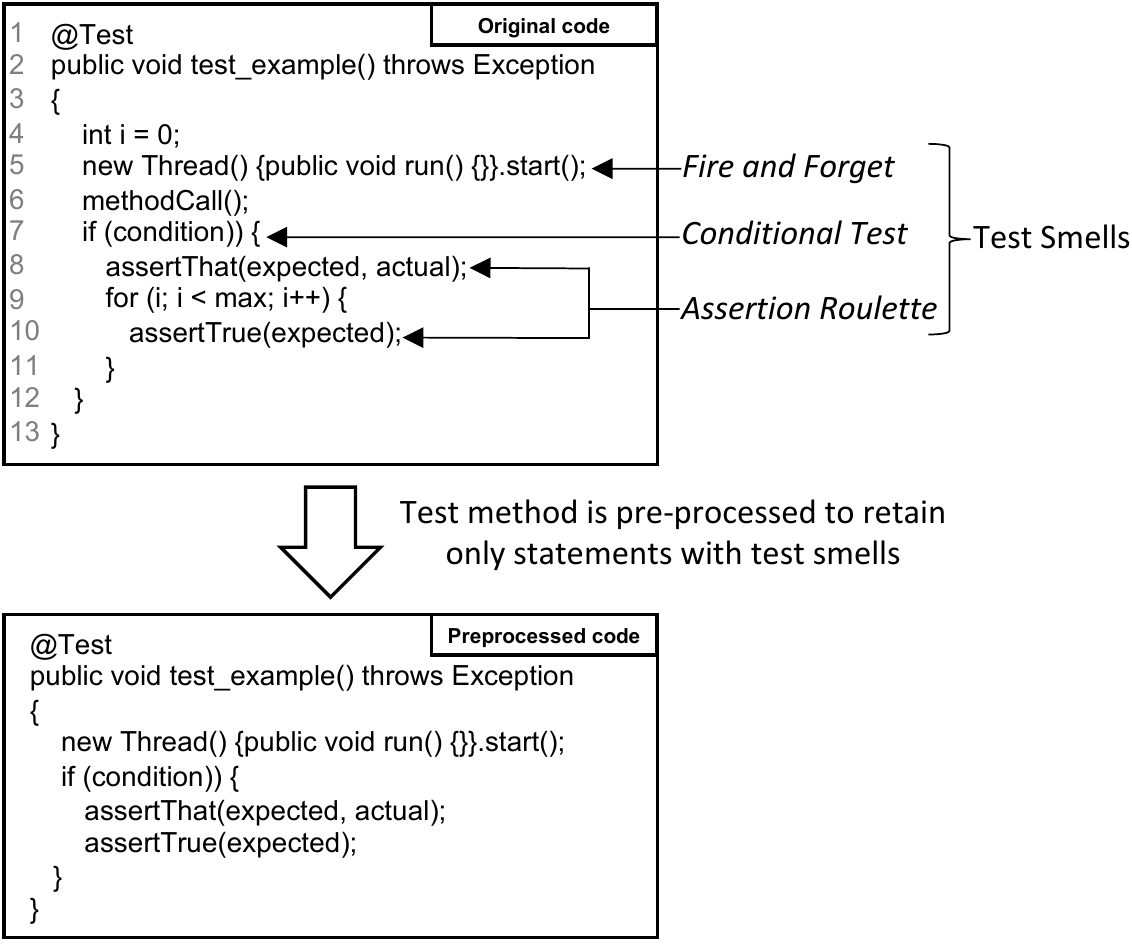}
        \vspace{-10pt}
        \caption{Example of pre-processing the source code of a test case, which leads to reducing the number of tokens from 62 down to 43}
        \label{fig:code_preprocessing}
        \vspace{-12pt}
    \end{figure}

    Figure~\ref{fig:code_preprocessing} gives an example of a Java test method, \texttt{test\_example}, and how it is pre-processed. As we can see, \texttt{test\_example} has seven different statements, four of them having test smells. In particular, \texttt{test\_example} contains the following test smells: \textit{Fire and Forget} (line 5 -- launching a thread), \textit{Conditional Test} (line 7 -- \texttt{if} condition), and \textit{Assertion Roulette} (lines 8 and 10 -- multiple assertions). As a result, our technique retains only these four statements, which in turn leads to reducing the token length from 62 to 43 (31\% reduction rate). We expect our test code pre-processing to help improve the classification performance, since it mitigates the random truncation of code statements induced by CodeBERT.
    
\subsubsection{Heuristics for detecting test smells}
To detect test smells in test code, we followed the same detection heuristics as those used by Alshammari et al.~\cite{alshammari2021flakeflagger}. However, different from this work, which extracts test smell information dynamically from the test and production code (code coverage), we detected test smells statically by analyzing the test code only. To this end, we used an Abstract Syntax Tree (AST)~\cite{noonan1985algorithm} parser, provided by the Eclipse JDT library,\footnote{\url{https://www.eclipse.org/jdt}} to traverse any given test code and retain statements that match, according to our heuristics, any of the targeted test smells. Using this library, each Java test file in the test suite is parsed and converted into AST nodes representing different code elements, e.g., method declaration or invocation. While parsing Java test files, not all types are necessarily resolved due to missing production code. We describe below the heuristics used to identify each of the eight test smells presented in Table~\ref{tab:TestSmells}. For each test case, i.e., test method, we analyzed each statement to check whether it matches one of the targeted test smells. If so, we retain that statement as part of the pre-processed test code and otherwise exclude that statement. For some test smells, we added flags, i.e., a Java line comments appended to the end of each statement matching the test smell, to help our fine-tuned model learn about the association of these statements with test flakiness. The test smells used in this work were detected as described below.

\vspace{-5pt}
\begin{itemize}
    \item \textbf{Indirect Testing:}
    We check whether a statement invokes a method that belongs to a class other than the test class or the production class under test. Since our approach is black-box, i.e., no access to production code, the production class name is extracted from the test class name by removing the word `\texttt{Test}'. This is a commonly used coding convention, but our approach can easily be adapted to other coding conventions in practice~\cite{palomba2017notice}. Any statement that is found to invoke such methods is retained and the `\texttt{//IT}' flag is added.
    
    \item \textbf{Eager Testing:}
    We check whether a test case invokes more than one method belonging to the production class under test as it can introduce confusion to what exactly a test method is testing~\cite{aljedaani2021test}. If this is the case, we retain the statements invoking these methods, adding the `\texttt{//ET}' flag.
    
    \item \textbf{Test Run War:}
    We check whether a statement accesses \texttt{static} variables that are not declared as \textit{final}, as the value of such variables could be changed by other test cases in different test executions, especially when a test case is order-dependent, which can then cause resource interference during test case execution~\cite{palomba2017notice}. Any statement that is found to use one of these variables is retained, adding the `\texttt{//RW}' flag.

    \item \textbf{Conditional Logic:}
    We check whether a statement contains an \texttt{if} condition. If so, we retain \texttt{if} statements, including their logical expressions. The presence of conditional statements makes test case behavior dependent on their logical expressions, thus making them unpredictable~\cite{aljedaani2021test}. For the statements inside the \texttt{then} and \texttt{else} blocks, we only retain those that match one of the eight test smells.
    
    \item \textbf{Fire and Forget:}
    We check whether a statement invokes a method that launches a thread by checking if the invoked method belongs to the \texttt{java.lang.Thread} class, \texttt{java.lang.Runnable} interface, or \texttt{java.util.concurrent} package. Thread-related statements make test cases prone to synchronization issues during their execution~\cite{pontillo2021toward}. If this test smell is present, we retain that statement.
    
    \item \textbf{Mystery Guest:}
    We check whether a statement invokes a method that accesses external resources, such as the file system (via \texttt{java.io.File}), database system (via \texttt{java.sql}, \texttt{javax.sql}, or \texttt{javax.persistence}), or network (via \texttt{java.net} or \texttt{javax.net}). Such external resources can introduce stability and performance issues during test case execution~\cite{pontillo2021toward}. Any statement that is found to use methods that belong to one of these classes or packages is retained. 
    
    \item \textbf{Assertion Roulette:}
     We check whether a statement performs one of the following assertion mechanisms, including \texttt{assertArrayEquals}, \texttt{assertEquals}, \texttt{assertFalse}, \texttt{assertNotNull}, \texttt{assertNotSame}, \texttt{assertNull}, \texttt{assertSame}, \texttt{assertThat}, \texttt{assertTrue}, and \texttt{fail}. If so, the statement is retained. Multiple assert statements in a test method makes it difficult to identify the cause of the failure if just one of the asserts fails~\cite{camara2021use}.
     
    \item \textbf{Resource Optimism:}
    We check whether a statement accesses the file system (\texttt{java.io.File}) without checking if the path (for either a file or directory) exists. Doing so makes optimistic assumptions about the availability of resources, thus causing non-deterministic behavior of the test case~\cite{peruma2020tsdetect}. We check the test initialization method (usually named as \texttt{setUp} or containing the \texttt{@Before} annotation) for any path checking method, including \texttt{getPath()}, \texttt{getAbsolutePath()}, or \texttt{getCanonicalPath()}. If no such checking is present, the statement is retained, adding the `\texttt{//RO}' flag.
\end{itemize}

\section{Validation}
\label{validation}
This section reports on the experiments we conducted to evaluate how accurate is \Flakify~in predicting flaky test cases and how it compares to FlakeFlagger as a baseline. We discuss the research questions we address, the datasets used, and the experiment design. Then, we present the results for each research question and discuss their practical implications.

\subsection{Research Questions}
        \begin{itemize}
            \item \textbf{RQ1: How accurately can \Flakify~predict flaky test cases?}
            
            The performance of ML-based flaky test predictors can be influenced by the data used for training and the underlying modeling methodology. In this RQ, we evaluate \Flakify~on two distinct datasets, which differ in terms of numbers of projects, ratios of flaky and non-flaky test cases, and the way flaky test cases were detected. In addition, predicting flaky test cases can be influenced by project-specific information used during model training, which is not available for new projects. Therefore, we evaluate \Flakify~using two different procedures: 10-fold cross-validation and per-project validation. The former mixes test cases from all projects together to perform model training and testing, whereas the later tests the model on every project such that no information from that project was used as part of model training.
            
            \vspace{3pt}
            \item \textbf{RQ2: How does \Flakify~compare to the state-of-the-art predictors for flaky test cases?}

            Many solutions have been proposed to predict flaky test cases. In this RQ, we compare the performance of our best performing model of \Flakify~(with test case pre-processing) to two versions (white-box and black-box) of FlakeFlagger, the best flaky test case predictor to date.
            
            \vspace{3pt}
            \textbf{\textit{RQ2.1: How accurate is \Flakify~for flaky test case prediction compared to the best white-box ML-based solution?}}
            \vspace{3pt}
            White-box prediction of flaky test cases requires access to production code, which is not (easily) accessible by software test engineers in many contexts. We assess whether \Flakify~achieves results that are at least comparable to the best white-box flaky test case predictor. Specifically, we compare the accuracy of the best performing model of \Flakify~with FlakeFlagger~\cite{alshammari2021flakeflagger}, the best white-box solution currently available, on the dataset used by FlakeFlagger. Our motivation is to determine whether black-box solutions, based on CodeBERT, can compete with the state-of-the-art, white-box ones. We compare the results of \Flakify~and FlakeFlagger on the dataset on which FlakeFlagger was evaluated, hereafter referred to as the FlakeFlagger dataset. We also performed a per-project validation of \Flakify~compared against FlakeFlagger to assess their relative capability to predict test cases in new projects.
            
            \vspace{3pt}
            \textbf{\textit{RQ2.2: How accurate is \Flakify~for black-box flaky test case prediction compared to the best ML-based solution?}}
            \vspace{3pt}
            Existing black-box flaky test case prediction solutions rely on a limited set of features that are sometimes project-specific or applicable only to a certain programming language, e.g., Java~\cite{pinto2020vocabulary}, since they were trained on features capturing the keywords of that language. Besides not being generic, the accuracy of these solutions has shown to be very low compared to white-box  solutions~\cite{alshammari2021flakeflagger}. Therefore, we compare the accuracy of \Flakify~with a black-box version of FlakeFlagger, by excluding the features related to production code, such as code coverage features (see Table~\ref{tab:FlakeFlagger_Features}). 

            \vspace{3pt}
            \item \textbf{RQ3: How does test case pre-processing improve \Flakify?}
            \vspace{3pt}
            The token length limitation of CodeBERT may lead to unintentionally removing relevant information about flaky test cases, which could then impact prediction accuracy. We assess whether the accuracy of \Flakify~is improved when training the model using pre-processed test cases containing only code statements related to test smells, as opposed to the entire test case code.  We fully realize that we may be missing test smells or unintentionally removing relevant statements. But our motivation is to assess the benefits, if any, of our approach to reduce the number of tokens used as input to CodeBERT. We performed this analysis on both the FlakeFlagger and the IDoFT datasets.
        \end{itemize}
     \begin{table*}
        \centering
        \caption{FlakeFlagger Features}
        \vspace{-7pt}
            \begin{tabular}{c|ll}
            \toprule
               \textbf{Category} & \textbf{Feature} &  \textbf{Description}\\
            \midrule
                \multirow{5}{*}{\rotatebox[origin=c]{90}{Black-Box}} 
                & Presence of Test Smells   &  See Table~\ref{tab:TestSmells} \\ 
                & Test Lines of Code        & Number of lines of code in the body of the test method \\
                & Number of Assertions      & Number of assertions checked by the test \\
                & Execution Time            & Running time for the test execution \\
                & Libraries                 & Number of external libraries used by the test \\
            \midrule
                \multirow{4}{*}{\rotatebox[origin=c]{90}{White-Box}} 
                & Source Covered Classes    & Number of production classes covered by each test \\
                & Source Covered Lines      & Number of lines covered by the test, counting only production code \\
                & Covered Lines             & Number of lines of code covered by the test \\
                & Covered Lines Churn       & Churn of covered lines in past 5, 10, 25, 50, 75, 100, 500, and 10,000 commits\\
            \bottomrule
            \end{tabular}
        \label{tab:FlakeFlagger_Features}
        \vspace{-10pt}
    \end{table*}
    
\subsection{Datasets Collection and Processing}
\label{sssec:dataset}
    To evaluate \Flakify, we used two publicly available datasets for flaky test cases. The first dataset is the FlakeFlagger dataset~\cite{alshammari2021flakeflagger}. The second dataset is the International Dataset of Flaky Tests (IDoFT)\footnote{\url{https://mir.cs.illinois.edu/flakytests}}, which comprises many datasets for flaky test cases used by previous studies on flaky test case prediction~\cite{wei2021probabilistic,lam2020large,lam2020understanding,lam2020dependent,shi2019ifixflakies,lam2019idflakies}.
    
    \vspace{4pt}
    \noindent\textbf{FlakeFlagger dataset:} It is provided by Alshammari et al.~\cite{alshammari2021flakeflagger}, containing flakiness information about 22,236 test cases collected from 23 GitHub projects. These projects have different test suite sizes, ranging from 55 to 6,267 (with a median of 430) test cases per project. All projects in the FlakeFlagger dataset are written in Java and use Maven as a build system, and each test case is a Java test method. The dataset contains the source code of each test case and the corresponding features that were computed to train FlakeFlagger. Also, test cases in the dataset were assigned labels indicating whether they are \textit{Flaky} or \textit{Non-Flaky}, which were determined by executing each test case 10,000 times.
    
    When we analyzed the dataset, we identified 453 test cases with missing source code when intersecting test cases in a provided CSV file (called \textit{processed\_data}\footnote{\url{https://github.com/AlshammariA/FlakeFlagger/blob/main/flakiness-predicter/result/processed_data.csv}}) with those in a provided folder (called \textit{original\_tests}\footnote{\url{https://github.com/AlshammariA/FlakeFlagger/tree/main/flakiness-predicter/input_data/original_tests}}) containing their source code. In addition, we identified 122 test cases, in the \textit{original\_tests} folder, with empty source code, which we found out were not written in Java.\footnote{\url{https://github.com/AlshammariA/FlakeFlagger/pull/4}} Therefore, we excluded these test cases from our dataset, since they do not add any valuable information regarding our flakiness prediction evaluation. Nine of these test cases were labeled as flaky, three with missing source code and six with empty method body. After excluding test cases with missing and empty code, we obtained 21,661 test cases for our experiments. We compared \Flakify~and FlakeFlagger using this updated dataset. To pre-process the source code of the test cases (see Section~\ref{Identifying_Test_Smells}), we cloned the GitHub repository of each project and extracted the Java classes defining the methods of test cases.
    
    There are 802 test cases in the dataset that are labeled as \textit{Flaky} (with a median of 19 flaky test cases per project), whereas 20,859 test cases are \textit{Non-Flaky}. About 4\% of all test cases exceed the 512 limit of CodeBERT when converted into tokens, including 14\% of the flaky test cases.
    
    \vspace{4pt}
    \noindent\textbf{IDoFT dataset:} This dataset contains 3,742 \textit{Flaky} test cases from 314 different Java projects, and collected using different ways, i.e., different runtime environments with different numbers of runs to detect test flakiness. However, we were unable to obtain the test code of 474 test cases (from 2 projects) due to missing GitHub repositories or commits, leaving us with 3,268 \textit{Flaky} test cases from 312 projects. Given that the IDoFT dataset contains no test cases categorized as \textit{Non-Flaky}, we used the fixed versions of 1,263 flaky test cases, from 174 projects, to obtain non-flaky test cases, as recommended by the IDoFT maintainers\footnote{\url{https://github.com/TestingResearchIllinois/IDoFT/issues/566}}. To do so, we relied on the provided links to pull requests\footnote{\url{https://mir.cs.illinois.edu/flakytests/fixed.html}} used for fixing flaky test cases to collect the corresponding code changes. However, of the 1,263 fixed flaky test cases, we found only 594 flaky test cases, from 126 projects, in which the test case code is changed to fix test flakiness. Based on our analysis, the other flaky test cases were fixed in other ways, such as changing the order of test case execution, test configuration, or production code. Such flaky tests are out of the scope of this paper, since we consider only test cases whose test code was fixed, e.g., causes of flakiness related to test smells or other test characteristics. As a result, we added the 594 \textit{Non-Flaky} (fixed) tests to the 3,268 \textit{Flaky} test cases to end up with an updated dataset of 3,862 test cases. Limitations, regarding the causes of flakiness we could not detect, are discussed in Section~\ref{threats}. About 13\% of all test cases exceed the 512 limit of CodeBERT when converted into tokens.

    We made the updated datasets of FlakeFlagger and IDoFT, including their pre-processed test cases, publicly available in our replication package~\cite{our_replication_package}.

    \subsection{Experiment Design}

    \subsubsection{Baseline}
         We used the FlakeFlagger approach as a baseline against which we compare the results achieved by \Flakify. To this end, we reran the experiments conducted by Alshammari et al.~\cite{alshammari2021flakeflagger} to reproduce the prediction results of FlakeFlagger using their provided replication package.\footnote{\url{https://github.com/AlshammariA/FlakeFlagger}} FlakeFlagger was trained and tested using a combination of white-box and black-box features listed in Table~\ref{tab:FlakeFlagger_Features}. These features were selected based on their Information Gain (IG), i.e., only features having an IG $\geq$ 0.01 were selected for training. Besides reproducing the original results of FlakeFlagger, we also reran the experiments using black-box features only, which was done by excluding all features that required access to production code. Comparing \Flakify~with FlakeFlagger is performed on the FlakeFlagger dataset only, as running FlakeFlagger on the IDoFT dataset requires extracting features, both dynamic and static, needed to train FlakeFlagger. To do so, we must access the project’s production code and then successfully execute thousands of test cases across hundreds of project versions.
       
    \subsubsection{Training and Testing Prediction Models}
        Training and testing \Flakify~were conducted using two different procedures, performed independently on the two datasets describe above, as follows.
         
        \vspace{4pt}\noindent\textbf{1$^{st}$ Procedure (cross-validation):} In this procedure, we evaluated \Flakify~similarly to how FlakeFlagger was originally assessed. Specifically, we used a 10-fold stratified cross-validation to ensure our model is trained and tested in a valid and unbiased way. For that, we allocated 90\% of the test cases for training and 10\% for testing our model in each fold. However, different from FlakeFlagger, we employed 20\% of the training dataset as a validation dataset, which is required for fine-tuning CodeBERT. Using the validation dataset, we calculated the training and validation loss, which helped obtain optimal weights and stop the training early enough to avoid overfitting. 

        Given that both of the datasets we used are highly imbalanced---\textit{Flaky} test cases represent only 3.7\% of all test cases in the FlakeFlagger dataset and \textit{Non-Flaky} test cases represent only 15\% of the IDoFT dataset---we balanced \textit{Flaky} and \textit{Non-Flaky} test cases in the training and validation datasets of FlakeFlagger and IDoFT. Different from FlakeFlagger, which used the synthetic minority oversampling technique (SMOTE)~\cite{chawla2002smote}, we used random oversampling~\cite{branco2016survey}, which adds random copies of the minority class to the dataset. We were unable to use SMOTE, since it requires vector-based features, whereas our model takes the source code of test cases (text) as input~\cite{feng2020codebert,pan2021empirical}, as opposed to pre-defined features like FlakeFlagger. Similar to FlakeFlagger, we also performed our experiments using undersampling but this led to lower accuracy. We did not balance the testing dataset to ensure that our model is only tested on the actual set of test cases. This prevents overestimating the accuracy of the model and reflects real-world scenarios where flaky test cases are rarer than non-flaky test cases~\cite{alshammari2021flakeflagger}.
             
        \vspace{4pt}
        \noindent\textbf{2$^{nd}$ Procedure (per-project validation):} In this procedure, we evaluated \Flakify~in a way that yields more realistic results when we predict test cases on a new project, thus evaluating the generalizability of \Flakify~across projects. To do this, we performed a per-project validation of \Flakify~on both datasets. In particular, for every project in each dataset, we trained \Flakify~on the other projects and tested it on that project. This allowed us to evaluate how accurate \Flakify~is in predicting flaky test cases in one project without including any data from that project during training. We also performed this analysis for FlakeFlagger, on the FlakeFlagger dataset, for the sake of comparison.
       
    \subsubsection{Evaluation Metrics}
        To evaluate the performance of our approach, we used standard evaluation metrics for ML classifiers, including \textit{Precision} (the ability of a classification model to precisely predict flaky test cases), \textit{Recall} (the ability of a model to predict all flaky test cases), and the \textit{F1-Score} (the harmonic mean of precision and recall)~\cite{goutte2005probabilistic}. For the per-project validation of \Flakify, we computed the overall precision, recall, and F1-score using the prediction results of all projects in the FlakeFlagger and IDoFT datasets. We also computed these metrics individually for those projects that have both \textit{Flaky} and \textit{Non-Flaky} test cases, specifically 23 FlakeFlagger projects and 126 IDoFT projects, along with descriptive statistics, such as mean, median, min, max, 25\% and 75\% quantiles. We used Fisher's exact test~\cite{raymond1995exact} to assess how significant is the difference in proportions of correctly classified test cases between two independent experiments. Note that precision, recall, and F1-score are computed based on such proportions.
        
        In practice, test cases classified as \textit{Flaky} must be addressed by re-running them multiple times or by fixing the root causes of flakiness~\cite{ziftci2020flake,GoogleFlakyMitigate,GoogleFlakyCost}. Precisely predicting flakiness is therefore important as otherwise time and resources are wasted on re-running and attempting to debug many test cases that are believed to be flaky but are not~\cite{memon2017taming,parry2021survey}. According to our industry partner, Huawei Canada, and a Google technical report~\cite{GoogleFlakyCost}, each flaky test case has to be investigated and re-run by developers. Hence, when we multiply the number of predicted flaky test cases, we proportionally increase the resources associated with re-running and investigating such flaky test cases. Therefore, we assume that the wasted cost of unnecessarily re-running and debugging test cases is inversely proportional to precision:
        
        \begin{equation}
            Test\ Debugging\ Cost  \propto  1-Precision
        \end{equation}
        
        On the other hand, it is also important not to miss too many flaky test cases as otherwise time is bound to be wasted on futile attempts to find and fix non-existent bugs in the production code. Thus, we assume that the wasted cost of unnecessarily finding and fixing non-existent bugs in the production code is inversely proportional to recall:

        \begin{equation}
            Code\ Debugging\ Cost  \propto  1-Recall
        \end{equation}

        We acknowledge that the above metrics are surrogate measures for cost and that there are significant differences between individual flaky tests; however, they are reasonable and useful approximations on large test suites for the purpose of comparing classification techniques. We used FlakeFlagger as baseline to compute the reduction rate of test and code debugging costs, by dividing the difference in cost between \Flakify~and FlakeFlagger by the cost of FlakeFlagger.

    \subsection{Results}
        \subsubsection{RQ1 results}
        Table~\ref{tab:Flakify_FlakeFlagger_all_results} shows the prediction results (in terms of precision, recall, and F1-score) of \Flakify~using both the full and pre-processed test code from the FlakeFlagger and IDoFT datasets, based on cross-validation. Overall, \Flakify~achieved promising prediction results using both datasets, with a precision of 70\%, a recall of 90\%, and an F1-score of 79\% on the FlakeFlagger dataset, and \noindent a precision of 99\%, a recall of 96\%, and an F1-score of 98\% on the IDoFT dataset. The higher results achieved by \Flakify~on the IDoFT dataset over those achieved on the FlakeFlagger dataset is probably due to the fact that the IDoFT dataset contains many more flaky test cases than FlakeFlagger, which helped during model training. Moreover, the non-flaky test cases in the IDoFT dataset were labeled based on developer's fixes addressing the causes of flakiness in the test code, unlike the non-flaky test cases in the FlakeFlagger dataset whose labels were based on 10,000 runs performed by Alshammari et al.~\cite{alshammari2021flakeflagger}, which may not have been enough to fully expose test flakiness. This also helped during model training of \Flakify.

        Table~\ref{tab:per_project_summary_results} reports the per-project prediction results of \Flakify~on the FlakeFlagger dataset. Overall, as expected, \Flakify~achieved slightly lower precision (72\%), recall (85\%), and F1-score (73\%) than the cross-validation results on the FlakeFlagger dataset. Similarly, \Flakify~achieved slightly worse precision (91\%), recall (88\%), and F1-score (89\%) on the IDoFT dataset. Table~\ref{tab:per_project_individual_results} shows descriptive statistics for the per-project prediction results of \Flakify~for individual projects of the FlakeFlagger dataset (due to space limitations, we provide individual per-project prediction results of \Flakify~on the IDoFT dataset in our replication package~\cite{our_replication_package}). Our analysis of individual per-project prediction results revealed a high performance of \Flakify~on the majority of projects. This result suggests that \Flakify~helps build models that are generalizable across projects, thus making it applicable to new projects where no historical information about test flakiness exists. In short, \Flakify~is capable to learn about test flakiness through data collected from other projects to predict flaky test cases in new projects.

        \begin{table*}[!htpb]
            \renewcommand{\arraystretch}{1.2}
            \centering
            \caption{Results of \Flakify~(using full code and pre-processed) compared to FlakeFlagger (white-box and black-box versions)}
    		\vspace{-5pt}
                \begin{tabular}{p{2.2cm}p{3.2cm}p{2.8cm}rrr}
                \toprule
                    \textbf{Approach} & \textbf{Dataset} & \textbf{Model} & \textbf{Precision} & \textbf{Recall} & \textbf{F1-Score} \\ 
                \midrule
                    \multirow{4}{*}{\Flakify} 
                             & \multirow{2}{*}{FlakeFlagger dataset}     & Full code          & 65\%          & 85\%           & 74\% \\
                             &                                           & Pre-processed code & \textbf{70\%} & \textbf{90\%}  & \textbf{79\%} \\
                                                                         \cdashline{2-6}
                             & \multirow{2}{*}{IDoFT dataset}            & Full code          & 98\%          & 95\%           & 92\% \\
                             &                                           & Pre-processed code & \textbf{99\%} & \textbf{96\%}  & \textbf{98\%} \\
                             \hdashline
                    \multirow{2}{*}{FlakeFlagger} 
                             & \multirow{2}{*}{FlakeFlagger dataset} & White-box version  & 60\%          & 72\%           & 65\% \\
                             &                                       & Black-box version  & 21\%          & 52\%           & 30\% \\
                \bottomrule
                \end{tabular}
            \vspace{5pt}
            \label{tab:Flakify_FlakeFlagger_all_results}
        \end{table*}

        \begin{table*}[!htpb]
            \renewcommand{\arraystretch}{1}
            \centering
            \caption{Summary of the per-project prediction results of \Flakify~on the FlakeFlagger and IDoFT datasets}
    		\vspace{-5pt}
                \begin{tabular}{p{3.2cm}lrrrrrrr}
                \toprule
                \textbf{Dataset} & \textbf{Metric} & \textbf{~~~~~~Min} & \textbf{~~~~~~25\%} &  \textbf{~~~~Mean} & \textbf{~~Median}  & \textbf{~~~~~~75\%}  & \textbf{~~~~~~Max}\\ 
                \midrule
                \multirow{3}{*}{FlakeFlagger dataset} & Precision & 6\%  & 58\%  & \textbf{72\% }& 79\%  & 91\%  & 100\% \\
                                                      & Recall    & 1\%  & 87\%  & \textbf{85\%} & 95\%  & 100\% & 100\% \\
                                                      & F1-Score  & 2\%  & 63\%  & \textbf{73\%} & 83\%  & 94\%  & 100\% \\
                \midrule
                \multirow{3}{*}{IDoFT dataset}        & Precision & 66\% & 100\% & \textbf{91\%} & 100\% & 100\% & 100\% \\
                                                      & Recall    & 14\% & 94\%  & \textbf{88\%} & 100\% & 100\% & 100\% \\
                                                      & F1-Score  & 25\% & 95\%  & \textbf{89\%} & 100\% & 100\% & 100\% \\
                \bottomrule
                \end{tabular}
            \vspace{5pt}
            \label{tab:per_project_summary_results}
        \end{table*}
        
        \begin{table*}[!htpb]
            \renewcommand{\arraystretch}{1}
            \centering
            \small
            \caption{Results of the per-project prediction for \Flakify~and FlakeFlagger on the FlakeFlagger dataset. For every project, we trained models on all other projects and tested them on that project.}
                \vspace{-5pt}
                \begin{tabular}{p{3.5cm}rrrrrr}
                \toprule
                    \multirow{2}{*}{\textbf{Project}} & \multicolumn{2}{c}{\textbf{Precision}} & \multicolumn{2}{c}{\textbf{Recall}} & \multicolumn{2}{c}{\textbf{F1-Score}}\\
                    & \textbf{\Flakify} & \textbf{FlakeFlagger} & \textbf{\Flakify} & \textbf{FlakeFlagger} & \textbf{\Flakify} & \textbf{FlakeFlagger}\\
                \midrule
                    achilles               & 100\%  &    0\%	& 100\%  &    0\%  & 100\%   &   0\%    \\ 
                    activiti               &  80\%  &    2\%	&  90\%  &   94\%  &  85\%   &   4\%    \\ 
                    alluxio                &  99\%  &  100\%   & 100\%  &   13\%  &  99\%   &  24\%    \\ 
                    ambari                 &  75\%  &   39\%	&  95\%  &   61\%  &  84\%   &  47\%    \\ 
                    assertj-core           &  25\%  &    0\%	& 100\%  &    0\%  &  40\%   &   0\%    \\ 
                    commons-exec           &  25\%  &    0\%	& 100\%  &    0\%  &  40\%   &   0\%    \\ 
                    elastic-job-lite&  50\%  &    0\%	& 100\%  &    0\%  &  60\%   &   0\%    \\ 
                    handlebars.java        &  30\%  &    0\%	& 100\%  &    0\%  &  50\%   &   0\%    \\ 
                    hbase                  &  79\%  &   72\%	&  98\%  &   33\%  &  88\%   &  45\%    \\ 
                    hector                 & 100\%  &    0\%	&  93\%  &    0\%  &  96\%   &   0\%    \\ 
                    http-request           &  88\%  &    0\%	&  88\%  &    0\%  &  88\%   &   0\%    \\ 
                    httpcore               &  74\%  &    7\%	&  90\%  &    4\%  &  81\%   &   5\%    \\ 
                    incubator-dubbo        &   6\%  &    7\%	&  16\%  &   32\%  &   9\%   &  12\%    \\ 
                    java-websocket         &  95\%  &    0\%	&  95\%  &    0\%  &  95\%   &   0\%    \\ 
                    logback                &  85\%  &    0\%	&  81\%  &    0\%  &  83\%   &   0\%    \\ 
                    ninja                  & 100\%  &    0\%	& 100\%  &    0\%  & 100\%   &   0\%    \\ 
                    okhttp                 &  78\%  &  100\%   &  85\%  &    2\%  &  81\%   &   4\%    \\ 
                    orbit                  &  88\%  &    0\%	& 100\%  &    0\%  &  93\%   &   0\%    \\ 
                    spring-boot            &  40\%  &    9\%	&   1\%  &    3\%  &   2\%   &   4\%    \\ 
                    undertow               &  75\%  &    7\%	&  85\%  &   43\%  &  79\%   &  12\%    \\ 
                    wildfly                &  65\%  &    6\%	&  91\%  &   26\%  &  76\%   &  10\%    \\ 
                    wro4j                  &  88\%  &    1\%	& 100\%  &   19\%  &  94\%   &   3\%    \\ 
                    zxing                  & 100\%  &    0\%	&  50\%  &    0\%  &  66\%   &   0\%    \\
                \midrule
                    Overall           &  \textbf{72\%}	 &   15\%	 &  \textbf{85\%}  &	  14\%	 &  \textbf{73\%}	 &   7\%  \\
                \bottomrule
                \end{tabular}
            \label{tab:per_project_individual_results}
            \vspace{-6pt}
        \end{table*}
        
        \subsubsection{RQ2 results}
        Table~\ref{tab:Flakify_FlakeFlagger_all_results} presents the prediction results of \Flakify, using both full code and pre-processed test code, and FlakeFlagger, using both white-box and black-box versions, for the FlakeFlagger dataset.
        
        \vspace{4pt}
        \noindent\textbf{\textit{RQ2.1 results.}} For FlakeFlagger, we obtained results close to those reported in the original study, with a slight decrease in F1-score (1\%), which is likely due to removing test cases with missing test code. \Flakify~achieved much better results with a precision of 70\% ($+$10 pp), a recall of 90\% ($+$18 pp), and an F1-score of 79\% ($+$14 pp). These results clearly show that \Flakify, though being black-box and relying exclusively on test code, significantly surpasses FlakeFlagger in accurately predicting flaky test cases. Statistically, the proportion of correctly predicted test cases using \Flakify~is significantly higher than that obtained with FlakeFlagger (Fisher-exact p-value $<0.0001$). 

        The number of true positives obtained by FlakeFlagger was 574, whereas \Flakify~increased that number to 721. This indicates that \Flakify~can potentially reduce the test debugging cost by 10 pp, as defined above, when compared to FlakeFlagger (a reduction rate of 25\%). Similarly, \Flakify~reduces the number of false negatives to 81 from 227 with FlakeFlagger, thus decreasing the code debugging cost by 18 pp, as defined above (a reduction rate of 64\%).
         
        Table~\ref{tab:per_project_individual_results} shows the comparison of per-project prediction results between \Flakify~and FlakeFlagger. Overall, \Flakify~achieves a high accuracy, with a precision of 72\% ($+$57 pp)), a recall of 85\% ($+$71 pp)), and an F1-score of 73\% ($+$66 pp)), which, once again, significantly outperforms FlakeFlagger. Looking at the individual prediction results of the projects, we observe that the accuracy of \Flakify~is largely consistent across projects, with a few exceptions, whereas FlakeFlagger performed poorly on the majority of projects. Further, \Flakify~performs better than FlakeFlagger for almost all projects except two: \texttt{incubator-dubbo} and \texttt{spring-boot} where both techniques fare poorly.
        
        To understand the reasons behind such degraded performance for these two projects, we performed a hierarchical clustering of the 23 projects. We used different metrics that capture the characteristics of each project, such as the number of test cases, number of flaky test cases, and frequency of test smells in each project. However, our clustering results were inconclusive, thus revealing no significant differences between the two projects and the other projects. As reported by Alshammari et al.~\cite{alshammari2021flakeflagger}, each project can have distinct characteristics, e.g., environmental setup and testing paradigm, that make it difficult to develop a general-purpose flaky test case predictor. For example, the \texttt{spring-boot} project has the highest number of flaky test cases among all projects, representing 20\% of all flaky test cases in the dataset. This, in turn, can influence model training when the model was tested for \texttt{spring-boot}. In addition, the variation in prediction results can be a result of a possible mislabeling of test cases as \textit{Flaky} and \textit{Non-Flaky} in some projects, since some test cases may still exhibit flakiness behavior if executed more than 10,000 executions, for example. Finally, test flakiness can also occur due to the use of network APIs or dependency conflicts~\cite{parry2022surveying}, which were not taken into account when predicting flaky test cases.

        \vspace{4pt}
        \noindent\textbf{\textit{RQ2.2 results.}}
        As shown in Table~\ref{tab:Flakify_FlakeFlagger_all_results}, we observe a considerable decline in the accuracy for the black-box version of FlakeFlagger when compared to its original, white-box version, i.e., 39 pp less precise with a 54 pp decrease in F1-score. Specifically, black-box FlakeFlagger correctly predicted a significantly lower proportion of test cases than both \Flakify~and the original, white-box version of FlakeFlagger (Fisher-exact p-values $<0.0001$). As a possible explanation, based on the results of FlakeFlagger regarding the importance of features in predicting flaky test cases~\cite{alshammari2021flakeflagger}, the majority of features having high IG values were based on source code coverage. Hence, removing those features, to make FlakeFlagger black-box, is expected to significantly decrease its prediction power. The difference in accuracy between \Flakify~and the black-box version of FlakeFlagger is rather striking, with a large improvement of +49\% in F1-score (Fisher-exact p-value $<0.0001$). FlakeFlagger is therefore not a viable black-box option to predict flaky test cases. 
        
        \vspace{-1pt}
        \subsubsection{RQ3 results}
        With no code pre-processing, 898 (4\%) of the test cases of the FlakeFlagger dataset and 505 (13\%) of the test cases of the IDoFT dataset were truncated by CodeBERT to generate tokens of size 512. Such arbitrary code truncation is likely to affect how accurately \Flakify~can predict flaky test cases. Pre-processing test cases (see Section~\ref{Identifying_Test_Smells}) led to reducing the number of test cases being truncated to only 40 (from 898) in the FlakeFlagger dataset and 87 (from 505) in the IDoFT dataset, a large difference. As a result, we observe in Table~\ref{tab:Flakify_FlakeFlagger_all_results} that, with pre-processed test cases, \Flakify~predicted flaky test cases with 5 pp higher F1-score on the FlakeFlagger dataset and 6 pp higher F1-score on the IDoFT dataset. This corresponds to a significantly higher proportion of correctly predicted test cases (Fisher-exact p-value $=0.0008$) for the FlakeFlagger dataset. In practice, the impact of pre-processing is expected to vary depending on the token length distribution of test cases. This result suggests that retaining statements related to test smells in the test code contributed to making \Flakify~more accurate, which also confirms the association of test smells with flaky test cases reported by prior research~\cite{palomba2017notice}.
        
\subsection{Discussion}
\label{discussion}

\noindent\textbf{More accurate predictions with easily accessible information.}
Our results showed that our black-box prediction of flaky test cases performs significantly better than a white-box, state-of-the-art approach. This not only enables test engineers to predict flaky test cases without rerunning test cases, but also without accessing the production code of the system under test, a significant practical advantage in many contexts. The highest accuracy of our \Flakify~was achieved by only retaining relevant code statements matching eight test smells. Yet, there is still room for improvement in terms of accuracy, which could be achieved by retaining more relevant statements based on additional test smells. For example, retaining code statements related to other common flakiness causes~\cite{parry2021survey}, such as concurrency and randomness, could further improve flaky test case predictions. However, the more code statements we retain, the more tokens to be considered by CodeBERT, which might lead to many test cases exceeding their token length limit, thus truncating other useful information. Hence, retaining additional code statements is a trade-off and should carefully be performed in balance with the resulting token length of test cases. Moreover, building a white-box flaky test predictor, by considering both production and test code, is not always technically feasible, since the production code is not always available to test engineers and, when possible, code coverage can be expensive and not scalable on large systems, especially in a continuous integration context. Considering the production code also makes it impractical to build language model-based predictors for flaky test cases, given the token length limitation of language models in general, and CodeBERT in particular. Nevertheless, future research should assess the practicability of white-box, model-based flaky test prediction, and should investigate further code pre-processing methods to make the use of language models more applicable in practice.

\vspace{3pt}
\noindent\textbf{Practical implications of imperfect prediction results.}
Though \Flakify~surpassed the best state-of-the-art solution in predicting flaky test cases, both in terms of precision and recall, a precision of 70\% is still not satisfactory, since misclassifying non-flaky test cases as flaky leads to additional, unnecessary cost, e.g., attempting to fix the test cases incorrectly predicted as flaky. Also, with a recall of 90\%, we miss 10\% of flaky test cases, leading to wasted debugging cost. If we assume that precision should be prioritized over recall, we can increase the former by restricting flaky test case predictions to those test cases with highest prediction confidence, at the expense of a lower recall. For example, this can be achieved by adjusting the classification threshold for flaky test cases to 0.60 or 0.70, instead of the default threshold of 0.50. Nevertheless, given that the predicted probabilities generated by the neural network in \Flakify~are over confident due to the use of the \textit{Softmax} function in the last layer~\cite{melotti2020probabilistic}, i.e., probabilities are either close to $0.0$ or $1.0$, we were unable to perform such analysis. Therefore, future research should employ techniques for calibrating the predicted probabilities~\cite{guo2017calibration} and enable threshold adjustments when classifying flaky test cases.

\vspace{3pt}
\noindent\textbf{Deployment of a flaky test case predictor in practice.}
\Flakify~can be deployed in Continuous Integration (CI) environments to help detect flaky test cases. One could argue that the CI build history can be used as reference to conclude whether a test case is flaky or not. However, regular test case executions across builds may not entirely solve the problem, since differences in test case verdicts, i.e., pass or fail, can be due to differences in builds rather than flakiness. Therefore, test engineers can use the prediction results obtained from \Flakify~to fix test cases that are predicted as flaky, e.g., by eliminating the presence of test smells, or otherwise rerun them a larger number of times, using the same code version, to verify whether a test case is actually flaky or not. More specifically, \Flakify~helps test engineers focus their attention on a small subset of test cases that are most likely to be flaky in a CI build.
As our results show, \Flakify~significantly reduces the cost of debugging test and production code, both in terms of human effort and execution time. This makes \Flakify~an important strategy in practice to achieve scalability, especially when applied to large test suites.
Moreover, the test smell detection capability of \Flakify~helps to inform test engineers about possible causes of flakiness that need to be addressed.

\vspace{-4pt}
\section{Threats to Validity}
\label{threats}
This section discusses the potential threats to the validity of our reported results.

    \vspace{-4pt}
    \subsection{Construct Validity}
    Construct threats to validity are concerned with the degree to which our analyses measure what we claim to analyze. In our study, to pre-process test cases, we used heuristics to retain code statements that match at least one of the eight test smells shown in Table~\ref{tab:TestSmells}. However, our heuristics might have missed some code statements having test smells and this could have led to suboptimal results when applying our approach. To mitigate this issue, though our approach to identify test smells is entirely different, we relied on the same heuristics as those used by Alshammari et al.~\cite{alshammari2021flakeflagger}. These heuristics assume commonly used coding conventions that might not be followed in all test suites. For example, we assumed that the test class name contains the production class name with the word `\textit{Test}'. However, such heuristics can easily be adapted to other coding conventions in practice. We also manually checked a random sample of test cases to verify that pre-processed code contains, as expected, only test smells-related code statements and does not dismiss any of them. We have made the tool we developed to detect test smells publicly available in our replication package~\cite{our_replication_package}.

    \vspace{-4pt}
    \subsection{Internal Validity}
    Internal threats to validity are concerned with the ability to draw conclusions from our experimental results. In our study, we used CodeBERT to perform a binary classification of test cases as \textit{Flaky} or \textit{Non-Flaky}. However, due to the token length limit of CodeBERT, the source code of some test cases was truncated, possibly leading to discarding relevant information about test flakiness. To mitigate this issue, we pre-processed the source code of test cases to retain only code statements related to test smells. Doing so did not only reduce the token length of test cases, but also improved the prediction power of our approach. However, our pre-processing may not be perfect or complete as it can lead to losing other relevant information. Future research should investigate whether retaining additionally relevant information to flaky test cases leads to improving prediction results, e.g., statements related to common flakiness causes, such as synchronous or platform-dependent operations.
    
    Moreover, our prediction results were compared with those of FlakeFlagger. But FlakeFlagger used white-box features, whereas our approach is black-box and the comparison may not be entirely meaningful. To mitigate this issue, we also compared our results with a black-box version of FlakeFlagger in which we removed any features requiring access to production code. In both cases, our approach obtained significantly higher prediction results than FlakeFlagger. We did not compare our results with other black-box approaches, e.g., vocabulary-based~\cite{pinto2020vocabulary}, since they are project-specific and did not achieve good results on the FlakeFlagger dataset~\cite{alshammari2021flakeflagger}.

    Finally, in our analysis, the cost of debugging the production or testing code assumes that test engineers address all test cases predicted as flaky. However, test engineers may choose to ignore a flaky test case, either by removing or skipping it, thus not introducing any cost. Yet, we believe that every flaky test case should be carefully addressed by test engineers, since ignoring test cases can lead to other kinds of costs, such as overlooked system faults.
    
    \vspace{-4pt}
    \subsection{External Validity}
    External threats are concerned with the ability to generalize our results. Our study is based on data collected by Alshammari et al.~\cite{alshammari2021flakeflagger}, which was obtained by rerunning test cases 10,000 times. Such data is of course not perfect as some test cases that were not found to be flaky could have been if rerun more times. To mitigate this threat, we used the same dataset for comparing \Flakify~with the baseline approach, FlakeFlagger. We also filtered out test cases which, to our surprise, had no source code in the dataset. Further, the FlakeFlagger and IDoFT datasets contain test cases from projects that are exclusively written in Java, which might affect the generalizability of our results. To mitigate this issue, we used CodeBERT, which was trained on six programming languages. Hence, we believe our approach would be applicable to projects written in other programming languages as well, given an appropriate tool to identify test smells.
    
    Moreover, CodeBERT was pre-trained on production source code only, i.e., source code related to test suites was not part of pre-training, making it unable to recognize test-specific structure and vocabulary, e.g., assertions. This can potentially increase token length, since test-specific key terms are decomposed into multiple tokens instead of one. For example, CodeBERT converts \texttt{assertEquals} into three tokens: \texttt{assert}, \texttt{\#\#equal}, and \texttt{\#\#s}, rather than just one token. Our pre-processing of the source code of test cases helped to mitigate the issue of token length; yet, future work should aim at pre-training CodeBERT on test code in addition to production code.
    
    Finally, the IDoFT dataset has shown that a significant number of test cases are flaky due to reasons unrelated to the test code. In situations where this is common, this is obviously a limitation of any black-box approach like \Flakify~relying exclusively on test code. In our evaluation, we did not consider such flaky test cases, but rather those whose causes of flakiness were in the test code, which were confirmed and manually fixed by developers, and thus considered in this paper as non-flaky. This helped during model training of \Flakify~on this dataset, which resulted in a higher prediction accuracy than those on the FlakeFlagger dataset.

\section{Related Work}
\label{related_work}
Flaky test detection has been an active area of research where many techniques were proposed to detect flaky test cases~\cite{parry2021survey}. Overall, these techniques can be classified into two groups: dynamic techniques, which require executing test cases to determine whether they are flaky or not, and static techniques, which rely only on the source code of test cases or the system under test. In this section, we review the flaky test detection techniques while comparing and contrasting them to our approach.

\subsection{ML-based Flaky Test Case Prediction}
A common approach to detect flaky test cases is to re-run test cases multiple times~\cite{parry2021survey,zolfaghari2021root}, which is computationally expensive. To address this issue, recent research has proposed the use of ML techniques for predicting flaky test cases, enabling test engineers to re-run only those test cases that are predicted to be flaky, thus reducing the cost of unnecessary debugging of test cases or production code.

Alshammari et al.~\cite{alshammari2021flakeflagger} proposed an innovative approach to predict flaky test cases using dynamically computed features capturing code coverage, execution history, and test smells. They re-ran test cases 10,000 times to identify whether a test case was flaky or not and thus establish a ground truth. Their prediction model predicted flaky test cases with an F1-score of 0.65, leaving significant room for improvement. However, some of the significant features required access to production files which, as discussed above, are not always accessible by test engineers or may not be computable in a scalable way in many practical contexts. Further, when only black-box features (see Table~\ref{tab:FlakeFlagger_Features}) were used, the F1-score decreased by 35 pp. In contrast, our approach achieved more accurate prediction results, with an F1-score of 0.79, while using test code only, thus offering a favorable black-box alternative.
    
In addition, Pontillo et al.~\cite{pontillo2021toward} proposed an approach to identify the most important factors associated with flaky test cases using the iDFlakies dataset~\cite{lam2019idflakies}. They used logistic regression to model flaky test cases using features that were statically computed using production code, e.g., code coverage, and test code, e.g., test smells. They found that code complexity (both production and test code), assertions, and test smells are associated with test flakiness. 
    
Another approach was proposed by Pinto et al.~\cite{pinto2020vocabulary} in which Java keywords were extracted from test code and employed as vocabulary features to predict test flakiness. Further, their study relied on the dataset of DeFlaker~\cite{bell2018deflaker}, in which test cases were re-run less than 100 times to establish the ground truth. Despite high accuracy results (F1-score = 0.95) on their dataset, their approach achieved much worse results (F1-score = 0.19) when using the dataset provided by Alshammari et al.~\cite{alshammari2021flakeflagger}. In addition, their models were language- and project-specific, since most of the significant features for predicting flaky test cases were related to Java keywords, e.g., \textit{throws}, or specific variable names, e.g., \textit{id}.In contrast, while our approach relies exclusively on test code, it builds a generic model to predict flakiness, based on features that are neither language- nor project-dependent, and achieved much better prediction results when using the FlakeFlagger dataset used by Alshammari et al.~\cite{alshammari2021flakeflagger}.

Moreover, Haben et al.~\cite{haben2021replication} and Camara et al.~\cite{camara2021vocabulary} replicated the study by Pinto et al. using other datasets containing projects written in other programming languages, e.g., Python. They found that vocabulary-based approaches are not generalizable, especially when performing inter-project flaky test case predictions, since new vocabulary is needed for any new project or programming language. Haben et al. also showed that combining the vocabulary-based features with code coverage features does not significantly improve the prediction accuracy of such an approach.

In summary, unlike the ML-based approaches above, our approach is generic, black-box, and language model-based, thus not requiring access to production code or pre-definition of features. Instead, our approach relies solely on test code to predict whether a test case is flaky or not.

\vspace{-3pt}
\subsection{Flaky Test Case Prediction using Test Smells}
Camara et al.~\cite{camara2021use} proposed an approach for predicting test flakiness using test smells as prediction features. These features require access to the production code and can be extracted using tsDetect~\cite{peruma2020tsdetect}, a tool for detecting test smells, that was applied to the DeFlaker dataset~\cite{bell2018deflaker}. Their study yielded a relatively high prediction accuracy (F1-score = 0.83). Alshammari et al.~\cite{alshammari2021flakeflagger} also relied on test smells as part of their features for predicting flaky test cases. However, the information gain of test smell features tended to be much lower than code coverage features, suggesting they are less significant flaky test case predictors. In \Flakify, we also relied on the test smells used by Alshammari et al.~\cite{alshammari2021flakeflagger}. However, they were not used as features but to exclusively retain relevant test code statements for fine-tuning our CodeBERT model. Doing so improved the accuracy of \Flakify, thus reducing the cost of rerunning or debugging test cases.

\vspace{-3pt}
\subsection{Flaky Test Detection at Run Time}
Memon et al.~\cite{Zatko15} used a simple dynamic pattern matching approach to detect flaky test cases at \textsc{Google} by simply searching for certain textual patterns in test execution logs, e.g., \textit{pass-fail-pass}, to identify whether a test case is flaky or not. The accuracy of detecting flaky test cases using this approach was 90\%. Similarly, Kowalczyk et al.~\cite{kowalczyk2020modeling} detected flaky test cases at \textsc{Apple} by analyzing the behavior of test cases using two scores: \textit{Flip rate}, which measures the rate at which a test case alternates between \textit{pass} and \textit{fail}, and \textit{Entropy}, which quantifies the uncertainty of a test case. An aggregated value of these two scores was used to generate flakiness ranks for test cases, which were then used to represent test flakiness, distributed across the test cases in different services at \textsc{Apple}. This technique marked 44\% of test failures as flaky with less than 1\% loss in fault detection. The above approaches require test cases to be executed many times to determine whether they are flaky, which is often not practical for large industrial projects. Unlike these approaches, \Flakify~is able to predict flaky test cases without executing them, relying exclusively on test code. 

Bell et al.~\cite{bell2018deflaker} proposed DeFlaker, a tool for detecting flaky test cases using coverage information about code changes. In particular, a test case is labeled as flaky if it fails and does not cover any changed code. Out of 4,846 test failures, DeFlaker was able to label 39 pp of them as flaky, with a 95.5\% recall and a false positive rate of 1.5\%, outperforming the default way of detecting flaky test cases, i.e., by rerunning test cases using Maven~\cite{Wendelin2019}. Different from DeFlaker, Lam et al.~\cite{lam2019idflakies} proposed iDFlakies, which detects test flakiness by re-running test cases in random orders. This framework was used to construct a dataset containing 422 flaky test cases, with almost half of them being order-dependent.

The above approaches either depend on rerunning test cases multiple times, execution history (not available for new test cases), or production code, e.g., coverage information. In contrast, \Flakify~does not require repeated executions of test cases or any information about the production code, including code coverage.
 
\section{Conclusion}
\label{conclusion}
In this paper, we proposed \Flakify, a black-box solution for predicting flaky test cases using only the source code of test cases, as opposed to the system under test. Further, it does not require to rerun test cases multiple times and does not entail the definition of features for ML prediction. 

We used CodeBERT, a pre-trained language model, and fine-tuned it to classify test cases as flaky or not based exclusively on test source code. We evaluated our work on two distinct datasets, namely the FlakeFlagger and IDoFT datasets, using two different evaluation procedures: (1) cross-validation and (2) per-project validation, i.e., prediction on new projects. In addition, we pre-processed this source code by retaining only code statements that match eight test smells, which are expected to be associated with test flakiness. This aimed at addressing a limitation of CodeBERT (and related language models), which can only process 512 tokens per test case. We evaluated our approach in comparison with both white-box and black-box versions of FlakeFlagger, the best state-of-the-art, ML-based flaky test case predictor.  The main results of our study are summarized as follows:

\begin{itemize}
    \item \Flakify~achieves promising results on two different datasets (FlakeFlagger and IDoFT) and under two different evaluation procedures, one assuming \Flakify~predicts test cases from a new project and the other one simply relying on cross-validation.
    \item When predicting test cases in new projects, the accuracy of \Flakify~is slightly lower but still close to cross-validation results.
    \item With cross-validation, \Flakify~reduces by 10 pp and 18 pp of the cost bound to be wasted by the original, white-box version of FlakeFlagger due to unnecessarily debugging test cases and production code, respectively.
    \item Similar to cross-validation results, \Flakify~also significantly outperforms FlakeFlagger when predicting flaky test cases in new projects, for which the model was not trained.
    \item A black-box version of FlakeFlagger is not a viable option to predict flaky test cases as it is too inaccurate.
    \item When retaining only code statements related to test smells, \Flakify~predicted flaky test cases with 5 pp and 6 pp higher F1-score on the FlakeFlagger and IDoFT datasets, respectively.
\end{itemize}

Overall, existing public datasets \cite{alshammari2021flakeflagger,bell2018deflaker,lam2019idflakies,haben2021replication} are not fully adequate to appropriately evaluate flaky test case prediction approaches, since the ratio of flaky test cases tends to be very low. In addition, flaky test cases in these datasets were detected by rerunning test cases numerous times while monitoring their behavior across executions, a technique that may be inaccurate. Further, many open source projects nowadays adopt Continuous Integration (CI), which provides extensive test execution histories. Given the frequency of test executions in CI and the high workload on CI servers, test cases might expose further flakiness behaviors due to causes that may not be revealed when running test cases on machines dedicated to test execution~\cite{ghaleb2019studying,lampel2021life}. Therefore, we plan in the future to build a larger dataset of flaky test cases in a CI context.

Last, a significant proportion of flaky tests can be due to problems in the production code and cannot be addressed by black-box models. Therefore, in the future, we need to devise light-weight and scalable approaches to address such causes of flakiness. 

\section*{Acknowledgement}
This work was supported by a research grant from Huawei Technologies Canada, Mitacs Canada, as well as the Canada Research Chair and Discovery Grant programs of the Natural Sciences and Engineering Research Council of Canada (NSERC). The experiments conducted in this work were enabled in part by WestGrid (https://www.westgrid.ca) and Compute Canada (https://www.computecanada.ca). Moreover, we are grateful to the authors of FlakeFlagger and the maintainers of the IDoFT dataset, who have responded to our multiple inquiries for clarifications about the datasets. 

\bibliographystyle{unsrt}
\bibliography{IEEEabrv,references}

\begin{thebibliography}{10}

\bibitem{zolfaghari2021root}
Behrouz Zolfaghari, Reza~M Parizi, Gautam Srivastava, and Yoseph Hailemariam.
\newblock {Root causing, detecting, and fixing flaky tests: State of the art
  and future roadmap}.
\newblock {\em Software: Practice and Experience}, 51(5):851--867, 2021.

\bibitem{luo2014empirical}
Qingzhou Luo, Farah Hariri, Lamyaa Eloussi, and Darko Marinov.
\newblock {An empirical analysis of flaky tests}.
\newblock In {\em Proceedings of the 22nd ACM SIGSOFT International Symposium
  on Foundations of Software Engineering}, pages 643--653, 2014.

\bibitem{eck2019understanding}
Moritz Eck, Fabio Palomba, Marco Castelluccio, and Alberto Bacchelli.
\newblock {Understanding flaky tests: The developer’s perspective}.
\newblock In {\em Proceedings of the 2019 27th ACM Joint Meeting on European
  Software Engineering Conference and Symposium on the Foundations of Software
  Engineering}, pages 830--840, 2019.

\bibitem{bell2018deflaker}
Jonathan Bell, Owolabi Legunsen, Michael Hilton, Lamyaa Eloussi, Tifany Yung,
  and Darko Marinov.
\newblock {DeFlaker: Automatically detecting flaky tests}.
\newblock In {\em 2018 IEEE/ACM 40th International Conference on Software
  Engineering (ICSE)}, pages 433--444. IEEE, 2018.

\bibitem{lam2019idflakies}
Wing Lam, Reed Oei, August Shi, Darko Marinov, and Tao Xie.
\newblock {iDFlakies: A framework for detecting and partially classifying flaky
  tests}.
\newblock In {\em 2019 12th ieee conference on software testing, validation and
  verification (icst)}, pages 312--322. IEEE, 2019.

\bibitem{GoogleFlakyCost}
John Micco.
\newblock Advances in continuous integration testing at {Google}.
\newblock \url{https://research.google/pubs/pub46593}, 2018.

\bibitem{alshammari2021flakeflagger}
Abdulrahman Alshammari, Christopher Morris, Michael Hilton, and Jonathan Bell.
\newblock {FlakeFlagger: Predicting flakiness without rerunning tests}.
\newblock In {\em 2021 IEEE/ACM 43rd International Conference on Software
  Engineering (ICSE)}, pages 1572--1584. IEEE, 2021.

\bibitem{pinto2020vocabulary}
Gustavo Pinto, Breno Miranda, Supun Dissanayake, Marcelo d'Amorim, Christoph
  Treude, and Antonia Bertolino.
\newblock {What is the vocabulary of flaky tests?}
\newblock In {\em Proceedings of the 17th International Conference on Mining
  Software Repositories}, pages 492--502, 2020.

\bibitem{camara2021use}
Bruno Camara, Marco Silva, Andre Endo, and Silvia Vergilio.
\newblock {On the use of test smells for prediction of flaky tests}.
\newblock In {\em Brazilian Symposium on Systematic and Automated Software
  Testing}, pages 46--54, 2021.

\bibitem{feng2020codebert}
Zhangyin Feng, Daya Guo, Duyu Tang, Nan Duan, Xiaocheng Feng, Ming Gong, Linjun
  Shou, Bing Qin, Ting Liu, Daxin Jiang, and Ming Zhou.
\newblock {CodeBERT: A pre-trained model for programming and natural
  languages}.
\newblock {\em arXiv preprint arXiv:2002.08155}, 2020.

\bibitem{palomba2017notice}
Fabio Palomba and Andy Zaidman.
\newblock {Notice of retraction: Does refactoring of test smells induce fixing
  flaky tests?}
\newblock In {\em 2017 IEEE international conference on software maintenance
  and evolution (ICSME)}, pages 1--12. IEEE, 2017.

\bibitem{pontillo2021toward}
Valeria Pontillo, Fabio Palomba, and Filomena Ferrucci.
\newblock {Toward static test flakiness prediction: a feasibility study}.
\newblock In {\em Proceedings of the 5th International Workshop on Machine
  Learning Techniques for Software Quality Evolution}, pages 19--24, 2021.

\bibitem{ziftci2020flake}
Celal Ziftci and Diego Cavalcanti.
\newblock {De-flake your tests: Automatically locating root causes of flaky
  tests in code at google}.
\newblock In {\em 2020 IEEE International Conference on Software Maintenance
  and Evolution (ICSME)}, pages 736--745. IEEE, 2020.

\bibitem{lam2019root}
Wing Lam, Patrice Godefroid, Suman Nath, Anirudh Santhiar, and Suresh
  Thummalapenta.
\newblock {Root causing flaky tests in a large-scale industrial setting}.
\newblock In {\em Proceedings of the 28th ACM SIGSOFT International Symposium
  on Software Testing and Analysis}, pages 101--111, 2019.

\bibitem{bach2017coverage}
Thomas Bach, Artur Andrzejak, and Ralf Pannemans.
\newblock {Coverage-based reduction of test execution time: Lessons from a very
  large industrial project}.
\newblock In {\em 2017 IEEE International Conference on Software Testing,
  Verification and Validation Workshops (ICSTW)}, pages 3--12. IEEE, 2017.

\bibitem{haben2021replication}
Guillaume Haben, Sarra Habchi, Mike Papadakis, Maxime Cordy, and Yves Le~Traon.
\newblock {A Replication Study on the Usability of Code Vocabulary in
  Predicting Flaky Tests}.
\newblock In {\em 18th International Conference on Mining Software
  Repositories}, 2021.

\bibitem{parry2021survey}
Owain Parry, Gregory~M Kapfhammer, Michael Hilton, and Phil McMinn.
\newblock {A survey of flaky tests}.
\newblock {\em ACM Transactions on Software Engineering and Methodology
  (TOSEM)}, 31(1):1--74, 2021.

\bibitem{parry2022surveying}
Owain Parry, Gregory~M Kapfhammer, Michael Hilton, and Phil McMinn.
\newblock Surveying the developer experience of flaky tests.
\newblock In {\em Proceedings of the International Conference on Software
  Engineering: Software Engineering in Practice (ICSE-SEIP)}, 2022.

\bibitem{van2001refactoring}
Arie Van~Deursen, Leon Moonen, Alex Van Den~Bergh, and Gerard Kok.
\newblock {Refactoring test code}.
\newblock In {\em Proceedings of the 2nd international conference on extreme
  programming and flexible processes in software engineering (XP2001)}, pages
  92--95. Citeseer, 2001.

\bibitem{devlin2018bert}
Jacob Devlin, Ming-Wei Chang, Kenton Lee, and Kristina Toutanova.
\newblock {BERT: Pre-training of deep bidirectional transformers for language
  understanding}.
\newblock {\em arXiv preprint arXiv:1810.04805}, 2018.

\bibitem{peters2018deep}
Matthew~E Peters, Mark Neumann, Mohit Iyyer, Matt Gardner, Christopher Clark,
  Kenton Lee, and Luke Zettlemoyer.
\newblock {Deep contextualized word representations}.
\newblock {\em arXiv preprint arXiv:1802.05365}, 2018.

\bibitem{yang2019xlnet}
Zhilin Yang, Zihang Dai, Yiming Yang, Jaime Carbonell, Russ~R Salakhutdinov,
  and Quoc~V Le.
\newblock {XLNet: Generalized autoregressive pretraining for language
  understanding}.
\newblock {\em Advances in neural information processing systems}, 32, 2019.

\bibitem{liu2019roberta}
Yinhan Liu, Myle Ott, Naman Goyal, Jingfei Du, Mandar Joshi, Danqi Chen, Omer
  Levy, Mike Lewis, Luke Zettlemoyer, and Veselin Stoyanov.
\newblock {RoBERTa: A robustly optimized bert pretraining approach}.
\newblock {\em arXiv preprint arXiv:1907.11692}, 2019.

\bibitem{sun2019videobert}
Chen Sun, Austin Myers, Carl Vondrick, Kevin Murphy, and Cordelia Schmid.
\newblock {VideoBERT: A joint model for video and language representation
  learning}.
\newblock In {\em Proceedings of the IEEE/CVF International Conference on
  Computer Vision}, pages 7464--7473, 2019.

\bibitem{nadeau2007survey}
David Nadeau and Satoshi Sekine.
\newblock {A survey of named entity recognition and classification}.
\newblock {\em Lingvisticae Investigationes}, 30(1):3--26, 2007.

\bibitem{bach2007review}
Nguyen Bach and Sameer Badaskar.
\newblock {A review of relation extraction}.
\newblock {\em Literature review for Language and Statistics II}, 2:1--15,
  2007.

\bibitem{xu2019bert}
Hu~Xu, Bing Liu, Lei Shu, and Philip~S Yu.
\newblock {BERT post-training for review reading comprehension and aspect-based
  sentiment analysis}.
\newblock {\em arXiv preprint arXiv:1904.02232}, 2019.

\bibitem{sun2019fine}
Chi Sun, Xipeng Qiu, Yige Xu, and Xuanjing Huang.
\newblock {How to fine-tune BERT for text classification?}
\newblock In {\em China National Conference on Chinese Computational
  Linguistics}, pages 194--206. Springer, 2019.

\bibitem{vaswani2017attention}
Ashish Vaswani, Noam Shazeer, Niki Parmar, Jakob Uszkoreit, Llion Jones,
  Aidan~N Gomez, {\L}ukasz Kaiser, and Illia Polosukhin.
\newblock {Attention is all you need}.
\newblock In {\em Advances in neural information processing systems}, pages
  5998--6008, 2017.

\bibitem{mandic2001recurrent}
Danilo Mandic and Jonathon Chambers.
\newblock {\em {Recurrent neural networks for prediction: learning algorithms,
  architectures and stability}}.
\newblock Wiley, 2001.

\bibitem{schmidhuber1997long}
Sepp Hochreiter and J{\"u}rgen Schmidhuber.
\newblock {Long short-term memory}.
\newblock {\em Neural Comput}, 9(8):1735--1780, 1997.

\bibitem{keim2020does}
Jan Keim, Angelika Kaplan, Anne Koziolek, and Mehdi Mirakhorli.
\newblock {Does BERT Understand Code?--An Exploratory Study on the Detection of
  Architectural Tactics in Code}.
\newblock In {\em European Conference on Software Architecture}, pages
  220--228. Springer, 2020.

\bibitem{guo2020graphcodebert}
Daya Guo, Shuo Ren, Shuai Lu, Zhangyin Feng, Duyu Tang, Shujie Liu, Long Zhou,
  Nan Duan, Jian Yin, Daxin Jiang, and M.~Zhou.
\newblock {GraphCodeBERT: Pre-training code representations with data flow}.
\newblock {\em arXiv preprint arXiv:2009.08366}, 2020.

\bibitem{kanade2020learning}
Aditya Kanade, Petros Maniatis, Gogul Balakrishnan, and Kensen Shi.
\newblock {Learning and evaluating contextual embedding of source code}.
\newblock In {\em International Conference on Machine Learning}, pages
  5110--5121. PMLR, 2020.

\bibitem{jiang2021treebert}
Xue Jiang, Zhuoran Zheng, Chen Lyu, Liang Li, and Lei Lyu.
\newblock {TreeBERT: A tree-based pre-trained model for programming language}.
\newblock {\em arXiv preprint arXiv:2105.12485}, 2021.

\bibitem{husain2019codesearchnet}
Hamel Husain, Ho-Hsiang Wu, Tiferet Gazit, Miltiadis Allamanis, and Marc
  Brockschmidt.
\newblock {Codesearchnet challenge: Evaluating the state of semantic code
  search}.
\newblock {\em arXiv preprint arXiv:1909.09436}, 2019.

\bibitem{wu2016google}
Yonghui Wu, Mike Schuster, Z.~Chen, Quoc~V. Le, Mohammad Norouzi, Wolfgang
  Macherey, Maxim Krikun, Yuan Cao, Qin Gao, Klaus Macherey, Jeff Klingner,
  Apurva Shah, Melvin Johnson, Xiaobing Liu, Lukasz Kaiser, Stephan Gouws,
  Yoshikiyo Kato, Taku Kudo, Hideto Kazawa, Keith Stevens, George Kurian,
  Nishant Patil, Wei Wang, Cliff Young, Jason~R. Smith, Jason Riesa, Alex
  Rudnick, Oriol Vinyals, Gregory~S. Corrado, Macduff Hughes, and Jeffrey Dean.
\newblock {Google's neural machine translation system: Bridging the gap between
  human and machine translation}.
\newblock {\em arXiv preprint arXiv:1609.08144}, 2016.

\bibitem{clark2020electra}
Kevin Clark, Minh-Thang Luong, Quoc~V Le, and Christopher~D Manning.
\newblock {Electra: Pre-training text encoders as discriminators rather than
  generators}.
\newblock {\em arXiv preprint arXiv:2003.10555}, 2020.

\bibitem{pan2021empirical}
Cong Pan, Minyan Lu, and Biao Xu.
\newblock {An Empirical Study on Software Defect Prediction Using CodeBERT
  Model}.
\newblock {\em Applied Sciences}, 11(11):4793, 2021.

\bibitem{wu2021literature}
Jiajie Wu.
\newblock {Literature review on vulnerability detection using NLP technology}.
\newblock {\em arXiv preprint arXiv:2104.11230}, 2021.

\bibitem{howard2018universal}
Jeremy Howard and Sebastian Ruder.
\newblock {Universal language model fine-tuning for text classification}.
\newblock {\em arXiv preprint arXiv:1801.06146}, 2018.

\bibitem{agarap2018deep}
Abien~Fred Agarap.
\newblock {Deep learning using rectified linear units (ReLU)}.
\newblock {\em arXiv preprint arXiv:1803.08375}, 2018.

\bibitem{srivastava2014dropout}
Nitish Srivastava, Geoffrey Hinton, Alex Krizhevsky, Ilya Sutskever, and Ruslan
  Salakhutdinov.
\newblock {Dropout: a simple way to prevent neural networks from overfitting}.
\newblock {\em The journal of machine learning research}, 15(1):1929--1958,
  2014.

\bibitem{el2021bert}
Salma El~Anigri, Mohammed~Majid Himmi, and Abdelhak Mahmoudi.
\newblock {How BERT's dropout fine-tuning affects text classification?}
\newblock In {\em International Conference on Business Intelligence}, pages
  130--139. Springer, 2021.

\bibitem{yao2020adahessian}
Zhewei Yao, Amir Gholami, Sheng Shen, Mustafa Mustafa, Kurt Keutzer, and
  Michael~W Mahoney.
\newblock {ADAHESSIAN: An adaptive second order optimizer for machine
  learning}.
\newblock {\em arXiv preprint arXiv:2006.00719}, 2020.

\bibitem{aljedaani2021test}
Wajdi Aljedaani, Anthony Peruma, Ahmed Aljohani, Mazen Alotaibi, Mohamed~Wiem
  Mkaouer, Ali Ouni, Christian~D Newman, Abdullatif Ghallab, and Stephanie
  Ludi.
\newblock {Test smell detection tools: A systematic mapping study}.
\newblock {\em Evaluation and Assessment in Software Engineering}, pages
  170--180, 2021.

\bibitem{peruma2020tsdetect}
Anthony Peruma, Khalid Almalki, Christian~D Newman, Mohamed~Wiem Mkaouer, Ali
  Ouni, and Fabio Palomba.
\newblock {Tsdetect: An open source test smells detection tool}.
\newblock In {\em Proceedings of the 28th ACM Joint Meeting on European
  Software Engineering Conference and Symposium on the Foundations of Software
  Engineering}, pages 1650--1654, 2020.

\bibitem{virginio2020jnose}
T{\'a}ssio Virg{\'\i}nio, Luana Martins, Larissa Rocha, Railana Santana,
  Adriana Cruz, Heitor Costa, and Ivan Machado.
\newblock {JNose: Java Test Smell Detector}.
\newblock In {\em Proceedings of the 34th Brazilian Symposium on Software
  Engineering}, pages 564--569, 2020.

\bibitem{noonan1985algorithm}
Robert~E Noonan.
\newblock {An algorithm for generating abstract syntax trees}.
\newblock {\em Computer Languages}, 10(3-4):225--236, 1985.

\bibitem{wei2021probabilistic}
Anjiang Wei, Pu~Yi, Tao Xie, Darko Marinov, and Wing Lam.
\newblock {Probabilistic and systematic coverage of consecutive test-method
  pairs for detecting order-dependent flaky tests}.
\newblock In {\em International Conference on Tools and Algorithms for the
  Construction and Analysis of Systems}, pages 270--287. Springer, 2021.

\bibitem{lam2020large}
Wing Lam, Stefan Winter, Anjiang Wei, Tao Xie, Darko Marinov, and Jonathan
  Bell.
\newblock {A large-scale longitudinal study of flaky tests}.
\newblock {\em Proceedings of the ACM on Programming Languages},
  4(OOPSLA):1--29, 2020.

\bibitem{lam2020understanding}
Wing Lam, Stefan Winter, Angello Astorga, Victoria Stodden, and Darko Marinov.
\newblock {Understanding reproducibility and characteristics of flaky tests
  through test reruns in Java projects}.
\newblock In {\em 2020 IEEE 31st International Symposium on Software
  Reliability Engineering (ISSRE)}, pages 403--413. IEEE, 2020.

\bibitem{lam2020dependent}
Wing Lam, August Shi, Reed Oei, Sai Zhang, Michael~D Ernst, and Tao Xie.
\newblock {Dependent-test-aware regression testing techniques}.
\newblock In {\em Proceedings of the 29th ACM SIGSOFT International Symposium
  on Software Testing and Analysis}, pages 298--311, 2020.

\bibitem{shi2019ifixflakies}
August Shi, Wing Lam, Reed Oei, Tao Xie, and Darko Marinov.
\newblock {iFixFlakies: A framework for automatically fixing order-dependent
  flaky tests}.
\newblock In {\em Proceedings of the 2019 27th ACM Joint Meeting on European
  Software Engineering Conference and Symposium on the Foundations of Software
  Engineering}, pages 545--555, 2019.

\bibitem{our_replication_package}
{\Flakify: A Black-Box, Language Model-based Predictor for Flaky Tests --
  Replication Package}.
\newblock \url{https://doi.org/10.5281/zenodo.6994692}.

\bibitem{chawla2002smote}
Nitesh~V Chawla, Kevin~W Bowyer, Lawrence~O Hall, and W~Philip Kegelmeyer.
\newblock {SMOTE: synthetic minority over-sampling technique}.
\newblock {\em Journal of artificial intelligence research}, 16:321--357, 2002.

\bibitem{branco2016survey}
Paula Branco, Lu{\'\i}s Torgo, and Rita~P Ribeiro.
\newblock {A survey of predictive modeling on imbalanced domains}.
\newblock {\em ACM Computing Surveys (CSUR)}, 49(2):1--50, 2016.

\bibitem{goutte2005probabilistic}
Cyril Goutte and Eric Gaussier.
\newblock {A probabilistic interpretation of precision, recall and F-score,
  with implication for evaluation}.
\newblock In {\em European conference on information retrieval}, pages
  345--359. Springer, 2005.

\bibitem{raymond1995exact}
Michel Raymond and Fran{\c{c}}ois Rousset.
\newblock {An exact test for population differentiation}.
\newblock {\em Evolution}, pages 1280--1283, 1995.

\bibitem{GoogleFlakyMitigate}
John Micco.
\newblock Flaky tests at {Google} and how we mitigate them.
\newblock
  \url{https://testing.googleblog.com/2016/05/flaky-tests-at-google-and-how-we.html},
  2016.

\bibitem{memon2017taming}
Atif Memon, Zebao Gao, Bao Nguyen, Sanjeev Dhanda, Eric Nickell, Rob
  Siemborski, and John Micco.
\newblock {Taming Google-scale continuous testing}.
\newblock In {\em 2017 IEEE/ACM 39th International Conference on Software
  Engineering: Software Engineering in Practice Track (ICSE-SEIP)}, pages
  233--242, 2017.

\bibitem{melotti2020probabilistic}
Gledson Melotti, Cristiano Premebida, Jordan~J Bird, Diego~R Faria, and
  N~Gon{\c{c}}alves.
\newblock {Probabilistic Object Classification using CNN ML-MAP layers}.
\newblock {\em arXiv preprint arXiv:2005.14565}, 2020.

\bibitem{guo2017calibration}
Chuan Guo, Geoff Pleiss, Yu~Sun, and Kilian~Q Weinberger.
\newblock {On calibration of modern neural networks}.
\newblock In {\em International Conference on Machine Learning}, pages
  1321--1330. PMLR, 2017.

\bibitem{camara2021vocabulary}
Bruno Henrique~Pachulski Camara, Marco Aur{\'e}lio~Graciotto Silva,
  Andr{\'e}~Takeshi Endo, and Silvia~Regina Vergilio.
\newblock {What is the Vocabulary of Flaky Tests? An Extended Replication}.
\newblock In {\em 2021 2021 IEEE/ACM 29th International Conference on Program
  Comprehension (ICPC) (ICPC)}, pages 444--454, 2021.

\bibitem{Zatko15}
Atif Memon and John Micco.
\newblock How flaky tests in continuous integration.
\newblock \url{https://www.youtube.com/watch?v=CrzpkF1-VsA}, 2016.

\bibitem{kowalczyk2020modeling}
Emily Kowalczyk, Karan Nair, Zebao Gao, Leo Silberstein, Teng Long, and Atif
  Memon.
\newblock {Modeling and ranking flaky tests at Apple}.
\newblock In {\em 2020 IEEE/ACM 42nd International Conference on Software
  Engineering: Software Engineering in Practice (ICSE-SEIP)}, pages 110--119.
  IEEE, 2020.

\bibitem{Wendelin2019}
{Identifying and analyzing flaky tests in Maven and Gradle builds}.
\newblock \url{https://gradle.com/blog/flaky-tests}.
\newblock Accessed: 2021-11-01.

\bibitem{ghaleb2019studying}
Taher~Ahmed Ghaleb, Daniel~Alencar da~Costa, Ying Zou, and Ahmed~E Hassan.
\newblock {Studying the impact of noises in build breakage data}.
\newblock {\em IEEE Transactions on Software Engineering}, 47(09):1998--2011,
  2021.

\bibitem{lampel2021life}
Johannes Lampel, Sascha Just, Sven Apel, and Andreas Zeller.
\newblock {When life gives you oranges: detecting and diagnosing intermittent
  job failures at Mozilla}.
\newblock In {\em Proceedings of the 29th ACM Joint Meeting on European
  Software Engineering Conference and Symposium on the Foundations of Software
  Engineering}, pages 1381--1392, 2021.

\end{thebibliography}

\section*{Authors' Biographies}
\vspace{0pt}
\begingroup
\setlength{\intextsep}{-1.5pt}
\begin{wrapfigure}{l}{36mm} 
    \includegraphics[width=2.4in,height=1.87in,clip,keepaspectratio]{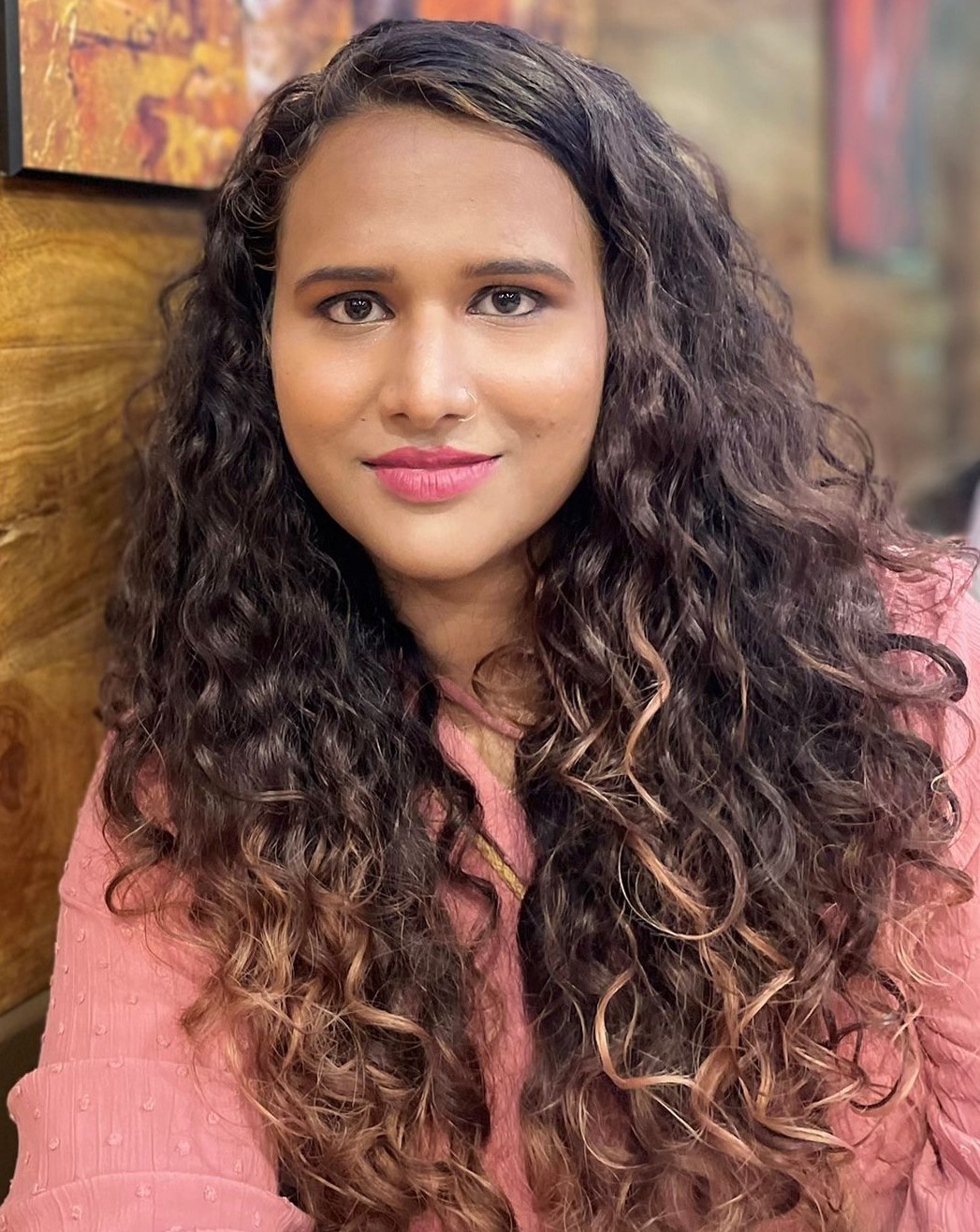}
  \end{wrapfigure}\par
  \textbf{Sakina Fatima} is a PhD candidate at the School of EECS at the University of Ottawa and a member of Nanda Lab. She obtained an Erasmus Mundus Joint Masters degree in Dependable Software Systems from the University of St Andrews, United Kingdom and Maynooth University, Ireland. In 2019, she was awarded the French Government Medal and the National University of Ireland prize for distinction in collaborative degrees. Her research interests include automated software testing and applied machine learning.\par
\endgroup

\vspace{8pt}

\begingroup
\setlength{\intextsep}{-1.5pt}
\begin{wrapfigure}{l}{36mm} 
    \includegraphics[width=2.3in,height=1.73in,clip,keepaspectratio]{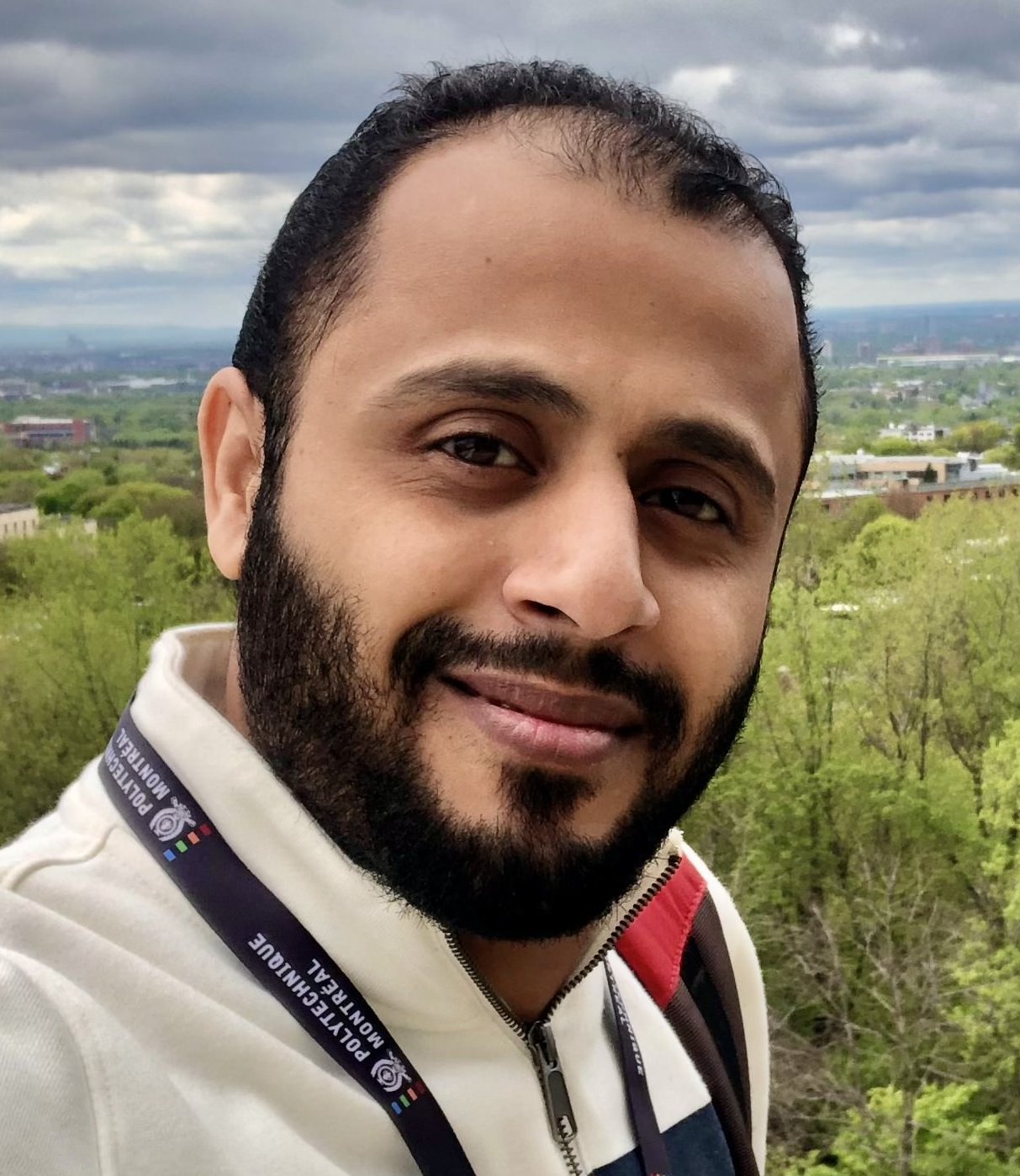}
  \end{wrapfigure}\par
  \noindent\textbf{Taher A. Ghaleb} is a Postdoctoral Research Fellow at the School of EECS at the University of Ottawa, Canada. Taher obtained his Ph.D. in Computing from Queen’s University, Canada (2021). During his Ph.D., Taher held an Ontario Trillium Scholarship, a highly prestigious award for doctoral students. He worked as a research/teaching assistant since he obtained his B.Sc. in Information Technology from Taiz University, Yemen (2008) and M.Sc. in Computer Science from King Fahd University of Petroleum and Minerals, Saudi Arabia (2016). His research interests include continuous integration, software testing, mining software repositories, applied data science and machine learning, program analysis, and empirical software engineering.\par
\endgroup

\vspace{8pt}

\begingroup
\setlength{\intextsep}{-1.5pt}
\begin{wrapfigure}{l}{36mm} 
    \includegraphics[width=2in,height=1.5in,clip,keepaspectratio]{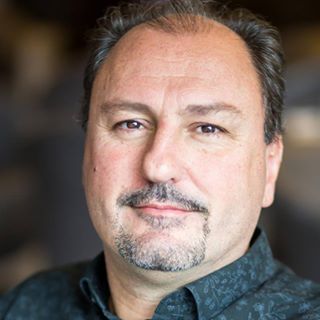}
  \end{wrapfigure}\par
  \noindent\textbf{Lionel C. Briand} is professor of software engineering and has shared appointments between (1) School of Electrical Engineering and Computer Science, University of Ottawa, Canada and (2) The SnT centre for Security, Reliability, and Trust, University of Luxembourg. He is the head of the SVV department at the SnT Centre and a Canada Research Chair in Intelligent Software Dependability and Compliance (Tier 1). He holds an ERC Advanced Grant, the most prestigious European individual research award, and has conducted applied research in collaboration with industry for more than 25 years, including projects in the automotive, aerospace, manufacturing, financial, and energy domains. He was elevated to the grades of IEEE and ACM fellow, granted the IEEE Computer Society Harlan Mills award (2012), the IEEE Reliability Society Engineer-of-the-year award (2013), and the ACM SIGSOFT Outstanding Research award for his work on software verification and testing. His research interests include: Testing and verification, search-based software engineering, model-driven development, requirements engineering, and empirical software engineering.\par
\endgroup
\end{document}